\documentclass{article}

\usepackage{arxiv}

\usepackage[utf8]{inputenc} 
\usepackage[T1]{fontenc}    
\usepackage{hyperref}       
\usepackage{url}            
\usepackage{booktabs}       
\usepackage{amsfonts}       
\usepackage{nicefrac}       
\usepackage{microtype}      
\usepackage{lipsum}		
\usepackage{graphicx}
\usepackage{natbib}
\usepackage{doi}

\title{ConsumerCheck: a software for analysis of sensory and consumer data}

\author{\href{https://orcid.org/0000-0003-1595-9962}{\includegraphics[scale=0.06]{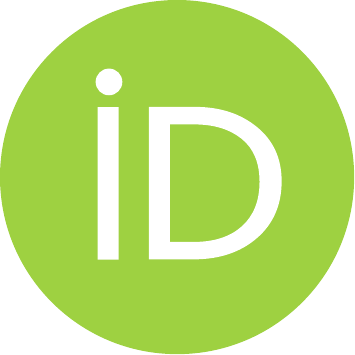}\hspace{1mm}Oliver Tomic}\thanks{https://olivertomic.wordpress.com/} \\
	Faculty of Science and Technology\\
	Norwegian University of Life Sciences\\
	\texttt{oliver.tomic@nmbu.no} \\
	\And
	\href{https://orcid.org/0000-0000-0000-0000}{\includegraphics[scale=0.06]{orcid.pdf}\hspace{1mm}Alexandra Kuznetsova} \\
	Leo Pharma\\
	Denmark\\
	\texttt{dkxdk@leo-pharma.com} \\
	\AND
	\href{https://orcid.org/00000-0002-1432-7229}{\includegraphics[scale=0.06]{orcid.pdf}\hspace{1mm}Per Bruun Brockhoff} \\
	Department of Applied Mathematics and Computer Science\\
	Technical University of Denmark\\
	\texttt{perbb@dtu.dk} \\
	\And
	Thomas Graff\\
	Faculty of Science and Technology\\
	Norwegian University of Life Sciences\\
	\texttt{thomas.graff@nmbu.no} \\
	\And
	\href{https://orcid.org/0000-0001-5610-3955}{\includegraphics[scale=0.06]{orcid.pdf}\hspace{1mm}Tormod N{\ae}s} \\
	Nofima\\
	Norway \\
	\texttt{tormod.naes@nofima.no} \\
}

\renewcommand{\headeright}{}
\renewcommand{\undertitle}{}
\renewcommand{\shorttitle}{ConsumerCheck software}

\hypersetup{
pdftitle={ConsumerCheck:A SOFTWARE FOR ANALYSIS OF SENSORY AND CONSUMER DATA},
pdfsubject={q-bio.NC, q-bio.QM},
pdfauthor={Oliver Tomic, Alexandra Kuznetsova, Per Bruun Brockhoff, Thomas Graff, Tormod Naes},
pdfkeywords={consumer liking data, descriptive analysis data, sensory profiling data, principal component analysis, PCA, preference mapping, partial least square regression, PLS, principal component regression, PCR, Conjoint analysis},
}

\begin{document}
\maketitle

\begin{abstract}
\textbf{ConsumerCheck} is an open source software for statistical analysis of data from sensory and consumer science. Such data are typically acquired from consumer trials and from descriptive analysis that was performed by trained sensory panels. ConsumerCheck comes with an easy-to-use graphical user interface that makes the implemented statistical methods easily accessible to users without programming skills, who are the majority of the target audience. In this way, ConsumerCheck meets the popular demand by users to get access to a domain specific data analysis software that provides a graphical user interface and that doesn't require a commercial license. Besides some simple descriptive statistics the main statistical methods implemented in ConsumerCheck are principal component analysis, preference mapping, partial least squares regression, principal component regression and conjoint analysis.
\end{abstract}


\section[Introduction]{Introduction}\label{sec:intro}
In sensory and consumer science \citep{lawless10} various types of methods exist for measurement of human responses triggered by sensory stimuli such as taste, smell, touch, etc. These methods generate data that require analysis with appropriate statistical methods \citep{naes10,martens01} in order to learn more about the consumers, their sensory preferences and their buying and consumption habits. Many of these statistical methods have been long available, either through commercial software, which come with user friendly graphical user interfaces (GUI) or as part of free statistical packages in open source programming languages such as \textbf{Python} or \textbf{R}. Relevant alternatives of these open source packages are the \textbf{Python} packages \textbf{scikit-learn} \citep{pedregosa11} and \textbf{statsmodels} \citep{seabold2010statsmodels} as well as \textbf{R} packages \textbf{SensMixed} \citep{SensMixed}, \textbf{sensR} \citep{sensR}, \textbf{SensoMineR} \citep{SensoMineR}, \textbf{chemometrics} \citep{filz15} and \textbf{pls} \citep{mevik15}. Some of the mentioned \textbf{R} packages may be used through a general GUI like \textbf{Rcmdr} \citep{fox15}. \\

In general, users without programming skills have two options when doing data analysis: (I) buy a commercial software with a GUI or; (II) learn how to program and use open source statistical packages to do the analysis at the command line or by writing code. Previous experience from similar research projects that produced some open source data analysis software \citep{tomic10PCH} has shown that most users in the target audience (i.e. research scientists, product developers of foods and beverages, lab personnel, etc.) would consider only software with a GUI since it makes the use of the statistical methods more easy and intuitive through a point-and-click approach. ConsumerCheck attempts to address this demand by providing a third option to the user: a domain specific open source software that comes with a GUI that is tailored towards each of the implemented statistical methods and as such makes them easy to apply. As one would expect from a GUI based software, ConsumerCheck also provides automated visualisation of the computational results through various specific plots. \\

Besides some simple descriptive statistics methods such as histograms and box plots for consumer liking data the main statistical methods implemented in ConsumerCheck are principal component analysis (PCA), preference mapping based on partial least squares regression (PLSR) and principal component regression (PCR), conjoint analysis as well as analysis of individual differences \citep{naes10}. One of the main concepts of ConsumerCheck is simplicity and ease of use, which is why only the most important statistical results are provided to the user. This means that the results provided by ConsumerCheck are not as exhaustive as those from the commercial software alternatives. The ConsumerCheck GUI is inspired by PanelCheck \citep{tomic09,tomic07}, a well established open source software \citep{tomic10PCH} within the field of sensometrics for performance analysis of trained sensory panels which has been available to the public in form of open source software since 2006.

The ConsumerCheck project home page (http://www.consumercheck.co/) provides important information about the software, links to downloads as well as well as installation instructions.

\section[Types of data and their properties]{Type of data and their properties}\label{sec:dataType}
ConsumerCheck was initially designed for analysis of four types of data that are common in sensory and consumer science, i.e. \textit{consumer liking} data, \textit{consumer characteristics} data, \textit{product design} data and \textit{descriptive analysis / sensory profiling data} (for more details see Section~\ref{sec:data_consumerLiking} through ~\ref{sec:data_sensory}). However, ConsumerCheck contains statistical methods that are generic, such as PCA, PLSR and PCR that allow for analysis of data data from any domain (see Section~\ref{sec:data_other}), not only from sensory and consumer science. The user only needs to keep in mind that the implemented statistical methods can analyse only data that are suitable for the method. For illustrational purposes five data sets from two sensory experiments, \textit{hams} and \textit{apples} respectively, will be used and analysed in this paper. More information on the \textit{hams} and \textit{apples} data are found in Section~\ref{sec:data_used}. It is important to note that as of ConsumerCheck version 1.3.3 all data (except for column and row names) need to be numerical. If factors in the conjoint analysis models (see Section~\ref{sec:statMethod_Conjoint} and ~\ref{sec:GUI_Conjoint}) are of type \textit{categorical} one should use numbers as factor levels instead of strings or characters. \\
Moreover, one needs to keep in mind that data with missing values may be imported, but at the current version of ConsumerCheck (2.3.1) none of the implemented statistical methods can handle them. When attempting to carry out computations with missing data an error message will be provided telling that computations are not possible. More information on how missing values are handled in ConsumerCheck are provided in Section~\ref{sec:GUI_DataImport_missingValues}.

\subsection[Consumer liking data]{Consumer liking data}\label{sec:data_consumerLiking}
\begin{figure}
  \centering
  \includegraphics[scale=0.6]{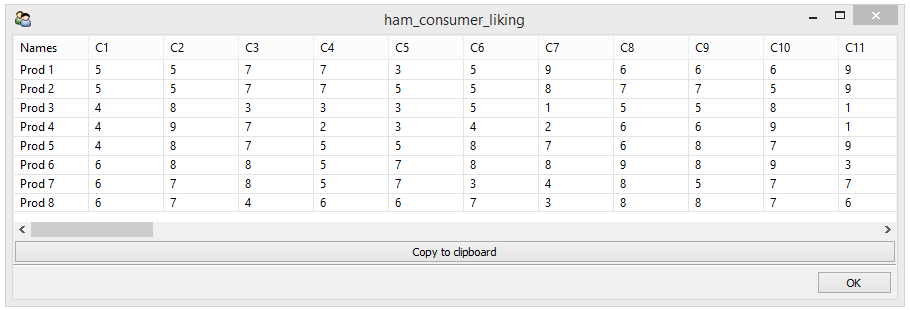}\\
  \caption{This is a screenshot showing a small part of the \textit{ham consumer liking} data that are described in Section~\ref{sec:data_used}. The data window shows product ratings of the first eleven consumers (C1 to C11) for the eight tested products.}\label{fig:Data_consumerLiking_ham}
\end{figure}

\textit{Consumer liking} data $X_{cl}$ are acquired through consumer trials where each consumer rates his or her liking of a product on a hedonic scale (typically from 1 to 5, 1 to 7 or 1 to 9, where the 1 represents "do not like at all" and the highest value represents "like very much"). The dimension of a \textit{consumer liking} data is $(J \times N)$ where the $j=1 \ldots J$ objects (products) are represented by rows and $n=1 \ldots N$ variables (consumers) are represented by columns. Figure~\ref{fig:Data_consumerLiking_ham} shows what \textit{consumer liking} data may look like.

Descriptive statistics and visualisation of the \textit{consumer liking} data distribution may be obtained with the methods implemented in \textit{Basic stat liking} (see Section~\ref{sec:statMethod_Basic}). Moreover, \textit{consumer liking} data can be analysed by using PCA (see Sections~\ref{sec:statMethod_PCA} and ~\ref{sec:GUI_PCA}); by using Preference mapping in combination with \textit{descriptive analysis / sensory profiling} data (see Sections~\ref{sec:statMethod_PrefMap} and \ref{sec:GUI_PrefMap}); by using PLSR/PCR in combination with either \textit{product design} data or \textit{consumer characteristics} data (see Section~\ref{sec:statMethod_PLSR}, \ref{sec:statMethod_PCR} and \ref{sec:GUI_PLSR_PCR}); by using conjoint analysis (see Section~\ref{sec:statMethod_Conjoint} and \ref{sec:GUI_Conjoint}) together with \textit{consumer characteristics} data and \textit{product design} data.

\subsection[Consumer characteristics data]{Consumer characteristics data}\label{sec:data_consumerCharacteristics}
\begin{figure}
  \centering
  \includegraphics[scale=0.6]{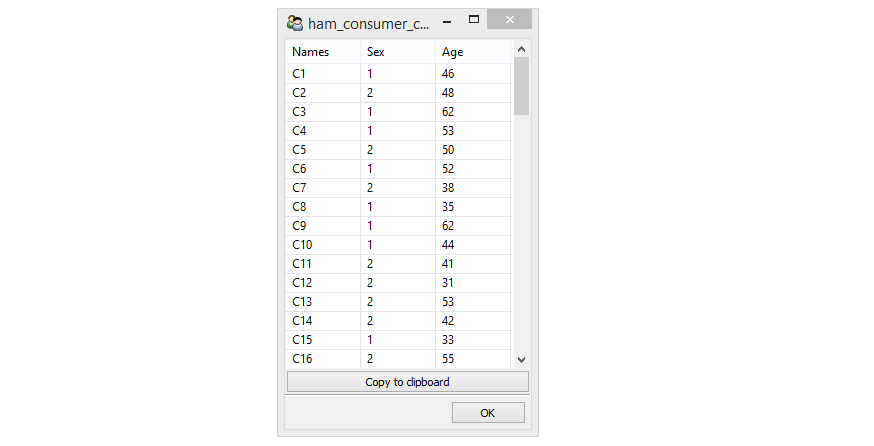}\\
  \caption{This is a screenshot showing a small part of the \textit{ham consumer characteristics} data that are described in Section~\ref{sec:data_used}. The data window shows two \textit{consumer characteristics} variables, i.e. sex and age of the first 16 consumers (C1 to C16).}\label{fig:Data_consumerCharacteristics_ham}
\end{figure}

\textit{Consumer characteristics} data $X_{cc}$ are data that provide background information on the consumers that have participated in the consumer trial. The \textit{consumer characteristics} data are of dimension $(N \times I)$ where the $n=1 \ldots N$ objects (consumers) are represented by rows and $i=1 \ldots I$ variables (consumer characteristics variables) are represented by columns. With the current version (1.3.3) of ConsumerCheck the consumer characteristics variables can be of any type, such as gender, age, country of origin, income, size of household, habits, etc as long as their levels are represented by integers. Note that the more levels a characteristics variable consists of the longer computation times will be. Although there is no limit to how many levels are allowed in a categorical variable we recommend to limit them to about five to six to keep computation times within reasonable limits. \textit{Consumer characteristics} data are usually analysed with conjoint analysis (see Section~\ref{sec:statMethod_Conjoint} and \ref{sec:GUI_Conjoint}) together with \textit{consumer liking} data and \textit{product design} data. There is no limit to how many characteristics variables the \textit{consumer characteristics} data may consist of, but one should not include more than three to four in the conjoint model, because computation time and complexity of the model would increase much. Instead, several models with fewer variables should be run to identify important characteristics. Figure~\ref{fig:Data_consumerCharacteristics_ham} shows what consumer characteristics data may look like.

\textit{Consumer characteristics} data can be analysed with PCA (see Sections~\ref{sec:statMethod_PCA} and ~\ref{sec:GUI_PCA}) provided that there are at least three variables. It is important to note that when analysing \textit{consumer characteristics} data all variables should be standardised since background variables are typically of different nature using different scales or units. Furthermore, it should be noted that it is not meaningful to include categorical variables (such as sex, where for example male is coded as 1 and female is coded as 2) and mix them with continuous variables (such as age) when applying PCA to the consumer characteristics data. Extended versions of PCA that handle this type of situation are available, such as the \textbf{R} package \textbf{PCAmixdata} \citep{chavent2014}, but this is not supported by ConsumerCheck at its current version.

\subsection[Product design data]{Product design data}\label{sec:data_design}
If the products rated by the consumers were produced by use of an experimental design, then these design data can be imported into ConsumerCheck and utilised for statistical analysis. The product design data $X_{d}$ are of dimension $(J \times M)$ where $j=1 \ldots J$ objects (products) are represented by rows and $m=1 \ldots M$ design variables are represented by columns.

\begin{figure}
  \centering
  \includegraphics[scale=0.6]{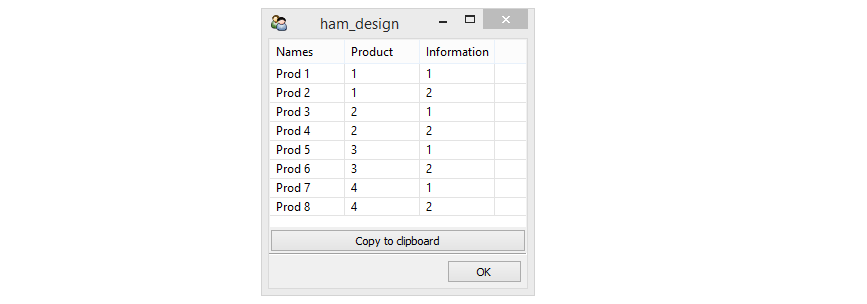}\\
  \caption{This is a screenshot showing the product design of the \textit{ham} data that are described in Section~\ref{sec:data_used}. The data window shows design variables \textit{Product} and \textit{Information} for the eight products. For more details on the \textit{ham} data see Section~\ref{sec:data_used}}\label{fig:Data_design_ham}
\end{figure}

The product design may be of type \textit{full factorial} or \textit{fractional factorial}. The product design data may be used for analysis in PLSR or PCR (Section~\ref{sec:statMethod_PLSR}, ~\ref{sec:statMethod_PCR} and ~\ref{sec:GUI_PLSR_PCR}) or in conjoint analysis (see Section~\ref{sec:statMethod_Conjoint} and \ref{sec:GUI_Conjoint}). Figure~\ref{fig:Data_design_ham} shows what product design data may look like.

\subsection[Descriptive analysis data]{Descriptive analysis or sensory profiling data }\label{sec:data_sensory}
\begin{figure}
  \centering
  \includegraphics[scale=0.6]{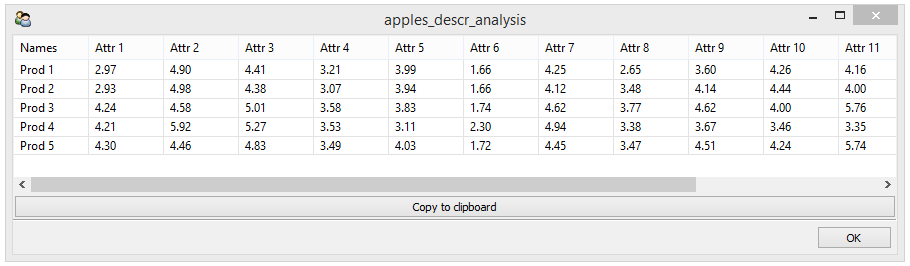}\\
  \caption{This is a screenshot showing a part of the \textit{apples descriptive analysis / sensory profiling} data that are described in Section~\ref{sec:data_used}. The data window shows intensity ratings for the first eleven sensory attributes for the five apple products that were tested by the trained sensory panel.}\label{fig:Data_descriptive_apple}
\end{figure}

Descriptive analysis (often also referred to as \textit{sensory profiling}) is a standard sensory tool that has an important role in research and product development \citep{lawless10}. When performing descriptive analysis a panel of trained assessors rate for each tested product the perceived intensity of defined sensory attributes on scales. The \textit{descriptive analysis / sensory profiling} data $X_{da}$ are of dimension $(J \times K)$ where the $j=1 \ldots J$ objects (food products) are represented by rows and $k=1 \ldots K$ variables (sensory attributes) are represented by columns. Figure~\ref{fig:Data_descriptive_apple} shows what \textit{descriptive analysis / sensory profiling} data may look like.

\textit{Descriptive analysis / sensory profiling} data can be analysed with PCA (see Sections~\ref{sec:statMethod_PCA} and ~\ref{sec:GUI_PCA}); with preference mapping (see Sections~\ref{sec:statMethod_PrefMap} and \ref{sec:GUI_PrefMap}) together with \textit{consumer liking} data; with PLSR or PCR (see Section~\ref{sec:statMethod_PLSR}, \ref{sec:statMethod_PCR} and \ref{sec:GUI_PLSR_PCR}) together with either \textit{product design} data or \textit{consumer characteristics} data.

\subsection[Relation between data types]{Relationship between the four types of sensory and consumer data}\label{sec:data_relation}
\begin{figure}
  \centering
  \includegraphics[scale=0.55]{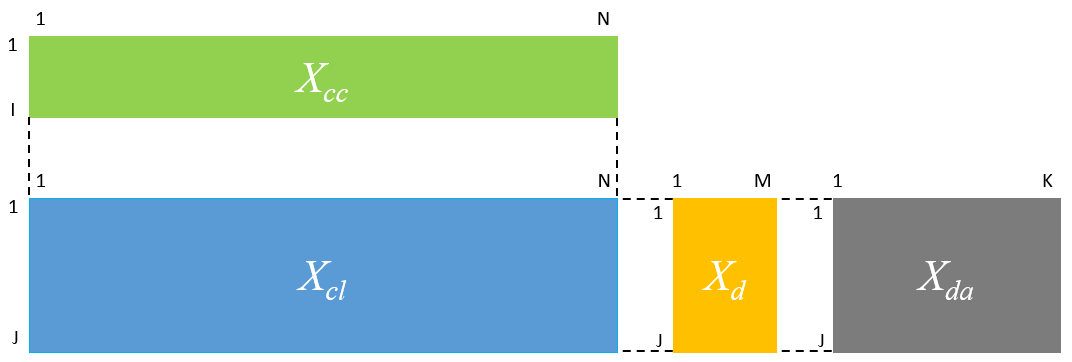}\\
  \caption{This plot illustrates how the four data types described above relate to each another. \textit{Consumer liking data $X_{cl}$}, \textit{product design matrix $X_{d}$} and \textit{descriptive analysis / sensory profiling data $X_{da}$} share the common axis or dimension of $J$ products. \textit{Consumer liking data $X_{cl}$} and \textit{consumer characteristics data $X_{cc}$} share the common axis or dimension of $N$ products.}\label{fig:L_shape}
\end{figure}

Figure~\ref{fig:L_shape} shows how the \textit{consumer liking} data $X_{cl}$, \textit{consumer characteristics} data $X_{cc}$, \textit{product design} matrix $X_{d}$ and \textit{descriptive analysis / sensory profiling} data $X_{da}$ relate to each other. Note that for illustration purposes the \textit{Consumer characteristics} data $X_{cc}$ are plotted transposed compared to how they are organised prior to import into ConsumerCheck. For $X_{cl}$, $X_{d}$ and $X_{da}$ the common axis are the tested products. For $X_{cl}$ and $X_{cc}$ the common axis are the consumers.

\subsection[Real world data used in examples]{Real world data used in examples}\label{sec:data_used}
In this paper a number of examples will illustrate how the statistical methods implemented in ConsumerCheck are applied to the data. For this purpose a couple of data sets are used that were acquired through two independent sensory and consumer science experiments. These two data sets are described in more detail below.

\subsubsection[Apple data]{Apple data}\label{sec:data_appleData}
The apple data consist of two data matrices: (I) a data matrix of type \textit{consumer liking} where 108 consumers have rated 5 apples. Hence its dimension is $(5 \times 108)$; (II) a data matrix of type \textit{descriptive analysis / sensory profiling} where the same 5 apples were described by a trained sensory panel using 14 attributes. Hence its dimension is $(5 \times 14)$.

\subsubsection[Ham data]{Ham data}\label{sec:data_hamData}
The ham data consist of three data matrices: (I) a data matrix of type \textit{consumer liking} (see Figure~\ref{fig:Data_consumerLiking_ham}) where 81 consumers rated the four hams twice (presented once as Norwegian and once as Spanish ham; see more details below in (III)). Hence its dimension of $(8 \times 81)$; (II) a data matrix of type \textit{consumer characteristics} (see Figure~\ref{fig:Data_consumerCharacteristics_ham}) consisting of two variables (named \textit{Sex} and \textit{Age}) that provide background information on the 81 consumers. Hence its dimension of $(81 \times 2)$; (III) a matrix of type \textit{product design} consisting of two design variables (see Figure~\ref{fig:Data_design_ham}). The first design variable is named \textit{Product} and represents the four hams that were presented the consumers to rate their liking. This means that there are four levels for design variable \textit{Product}. As part of the experiment each of the four hams were presented to the consumers twice, once pretending they were Norwegian ham and once pretending they were Spanish ham. The aim was to find out whether the country of produce would influence the liking of the consumer. This is determined by the second design variable, named \textit{Information}. It has two levels, where \textit{1} indicates that the ham was presented as Norwegian ham and \textit{2} indicates that it was presented as Spanish ham. From the two design variables we get a $(4 \times 2)$ full factorial experimental design which results in a total of eight ``unique'' ham products named \textit{Prod 1} through \textit{Prod 8}. The \textit{product design} data therefore is of dimension $(8 \times 2)$ where each row represent a unique combination of levels from the two design variables \textit{Product} and \textit{Information}.

\subsection[Other data types]{Other data types}\label{sec:data_other}
As mentioned above, ConsumerCheck was initially designed for analysis of data from sensory and consumer science (see Section~\ref{sec:data_consumerLiking} through ~\ref{sec:data_sensory}). However, since some statistical methods are generic there is no reason to limit the use of ConsumerCheck to only data from sensory and consumer science. Any kind of data that are suitable for analysis with PCA (Section~\ref{sec:statMethod_PCA}), PLSR (Section~\ref{sec:statMethod_PLSR}) and PCR (Section~\ref{sec:statMethod_PCR}) may be imported to ConsumerCheck and be tagged as type \textit{Other} when importing them.

\section[Statistical methods]{Statistical methods in ConsumerCheck}\label{sec:statMethod}
ConsumerCheck contains a number of statistical methods that are very common in analysis of sensory and consumer data. As of version 2.3.1 the following methods are implemented:
\begin{itemize}
  \item standard statistical methods for obtaining descriptive statistics from \textit{consumer liking} data such as box plots and histograms, see Section~\ref{sec:statMethod_Basic}
  \item principal component analysis (PCA), see Section~\ref{sec:statMethod_PCA}
  \item preference mapping (prefmap), see Section~\ref{sec:statMethod_PrefMap}
  \item partial least squares regression (PLSR), see Section~\ref{sec:statMethod_PLSR}
  \item principal component regression (PCR), see Section~\ref{sec:statMethod_PCR}
  \item conjoint analysis, see Section~\ref{sec:statMethod_Conjoint}
\end{itemize}

 All methods are thoroughly described in textbooks and scientific papers, which is why we keep the methods sections short and discuss in detail only issues that directly relate to the use of ConsumerCheck. Further on in this paper, the graphical user interface (GUI) for each method is discussed in detail (Section ~\ref{sec:GUI_Basic} through ~\ref{sec:GUI_Conjoint}), including how to set model parameters, how to obtain results and how plots and tables are interpreted.

\subsection[Basic statistics for consumer liking data]{Basic statistics for consumer liking data}\label{sec:statMethod_Basic}
Under the \textit{Basic stat liking} tab (see Figure~\ref{fig:GUI_basicStatLiking}) there are three types of plots available for quick acquisition of descriptive statistics from the \textit{consumer liking} data. The first two, that is the \textit{Box plot} and \textit{Stacked histogram}, visualise the distributions of the liking ratings across all consumers for each of the tested products or across all tested products for each consumer. The third type, the \textit{Single product histogram}, visualises the distribution of the ratings across all consumers for one specific product at the time in an ordinary histogram.

\subsubsection[Box plot]{Box plots}\label{sec:statMethod_Basic_box}
The \textit{box plot} (for an example see Figure~\ref{fig:GUI_basicStatLiking_boxplot}) describes how the consumer liking rates are distributed for each product. More precisely, it shows how the ratings are distributed between the 25 and 75 percentile of the data. The dark green line across the box indicates the median value. The vertical lines above and below a box indicate which range of the scale was used. More practical details on how to generate box plots in ConsumerCheck and the interpretation of results are found in Section~\ref{sec:GUI_Basic}.

\subsubsection[Stacked histogram]{Stacked histograms}\label{sec:statMethod_Basic_stacked}
Stacked histograms are another way of visualising the consumer liking rates for each product (see Figure~\ref{fig:GUI_basicStatLiking_stackedHistogram}). Here, however, one can see for every product how often each liking rate was used. More practical details on how to generate stacked histograms in ConsumerCheck and interpretation of results are found in Section~\ref{sec:GUI_Basic}.

\subsubsection[Histogram]{Single product histogram}\label{sec:statMethod_Basic_basicHisto}
These are ordinary histograms showing the distribution of ratings across all consumers for a single product. More practical details on how to generate histograms in ConsumerCheck and how to interpret results are provided in Section~\ref{sec:GUI_Basic}.

\subsection[Principal component analysis]{Principal component analysis}\label{sec:statMethod_PCA}
PCA \citep{mardia79} as implemented in ConsumerCheck is coded in \textbf{Python} and uses the NIPALS algorithm \citep{wold82} to provide scores, loadings, correlation loadings, calibrated and validated explained variances for the analysed data. Furthermore, one can access predicted (i.e. reconstructed) versions of the analysed data after each PC for both calibration and validation. PCA is accessible through the \textit{PCA} tab (see Figure~\ref{fig:GUI_PCA} and details on the usage is provided in Section~\ref{sec:GUI_PCA}). If needed, further computation results, such as root mean square error of calibration and cross validation (RMSEP and RMSECV, etc.), are available from the PCA class when using the \textbf{Python} source code directly outside ConsumerCheck. The PCA implemented in ConsumerCheck contains an option for variable standardisation if equal weight is to be given to each variable. It is important to note that ConsumerCheck automatically leaves out variables with zero variance when standardisation of variables is selected since the standard deviation for such a variable is STD=0. Whenever this happens, ConsumerCheck provides information on which variables have been left out in a message box dialog. The calibrated explained variance provided by the PCA model describes how much of the total variance in the data is explained by each principal component (PC). The cumulative calibrated explained variance (see Figure~\ref{fig:GUI_PCA_explainedVariance} for an example) increases with every PC added to the model. The validated explained variance is computed by systematically leaving out objects/rows from the data, then computing new PCA models and using the new loadings to predict values of the data that were left out. The closer the predictions of the left out data are to the real values of the left out data, the more robust the model. Note that the validated explained variance is computed using full cross validation, also known as leave-one-out in other scientific fields. Currently, for user friendliness and simplicity reasons there are no options to change this setting, but future versions of ConsumerCheck may provide k-fold cross validation. But for most of the practical cases in sensory and consumer analysis full cross validation should be sufficient, since the number of objects or products measured is usually low and the products typically independent. More detailed information on calibrated and validated explained variances are found elsewhere~\citep{martens89}.

\subsection[Preference mapping]{Preference mapping}\label{sec:statMethod_PrefMap}
Preference mapping \citep{green94, mcewan96} is a much used statistical method in the field of sensometrics that analyses \textit{consumer liking} and \textit{descriptive analysis / sensory profiling} data together. It is available through the \textit{Prefmap} tab. Preference mapping visualises individual differences between consumers and their preference for products with certain sensory attributes. The preference mapping model is actually a multivariate regression model that consists of an $X$ and $Y$ matrix and that attempts to find components that describe common variation between the two. Depending on whether the \textit{consumer liking} data or \textit{descriptive analysis / sensory profiling} data is chosen to be the $X$ matrix, one speaks of internal or external preference mapping ~\citep{naes10}, respectively. Furthermore, for the computation of the components one can choose between partial least square regression (PLSR) and principal component regression (PCR). Both PLSR (see Section~\ref{sec:statMethod_PLSR}) and PCR (see Section~\ref{sec:statMethod_PCR}) are well established multivariate regression methods in the field of sensometrics. Having the option to choose between the two can be seen as if these were two different "engines" that power the computations of the preference mapping model. The impact of making a choice between internal or external preference mapping combined with the selection between either PLSR and PCR is discussed elsewhere ~\citep{naes10}. Preference mapping and its engines PLSR and PCR are coded in \textbf{Python}. Details on the usage of preference mapping through its GUI is provided in Section~\ref{sec:GUI_PrefMap}. One can access $X$ scores, $X$ loadings, $Y$ loadings, calibrated and validated explained variances for $X$ and $Y$ as well as predictions of $Y$ after a number of components for calibration and validation. Using the \textbf{Python} source code one can access also results such as root mean square error of calibration and cross validation (RMSEP and RMSECV, etc.). As with PCA (see Section~\ref{sec:statMethod_PCA}), the calibrated explained variance is computed from the full set of objects/rows in $X$ an $Y$, whereas the validated explained variance is computed by use of full cross validation.

\subsection[Partial least squares regression]{Partial least squares regression}\label{sec:statMethod_PLSR}
PLSR~\citep{wold82} is a multivariate regression method that is frequently used in the field of sensometrics. The main purpose of the method is to find components that describe common variation between two data matrices $X$ and $Y$. In ConsumerCheck the NIPALS algorithm~\citep{wold82} is applied to compute results for PLSR. It searches for components by iterating forth and back between $X$ and $Y$, which means that both $X$ and $Y$ simultaneously influence the compuation of components unlike with PCR (see Section~\ref{sec:statMethod_PCR}) where only $X$ determines the components. \\
PLSR as implemented in ConsumerCheck is coded in \textbf{Python} and provides $X$ scores, $X$ loadings, $Y$ loadings, $X$ \& $Y$ correlation loadings, calibrated and validated explained variances for $X$ and $Y$ as well as predictions of $Y$ after a number of components for calibration and validation. If needed, further computation results, such as root mean square error of calibration (RMSEP) and cross validation (RMSECV), etc., are available from the PLSR class when using the \textbf{Python} source code directly outside ConsumerCheck. \\
PLSR is accessible through the \textit{Prefmap} tab (see Figure~\ref{fig:GUI_prefmap} and details on the usage in Section~\ref{sec:GUI_PrefMap}) and the \textit{PLSR/PCR} tab (see Figure~\ref{fig:GUI_PLSR-PCR} and details on the usage in Section~\ref{sec:GUI_PLSR_PCR}). Under the \textit{Prefmap} tab the use of PLSR is restricted to only \textit{consumer liking} data and \textit{descriptive analysis / sensory profiling} data, since \textit{preference mapping} deals only with these two types of data. Under the \textit{PLSR/PCR} tab other types of data may be analysed as for example \textit{consumer characteristics} data together with transposed \textit{consumer liking} data or \textit{product design} data together with \textit{consumer liking} data. If available, other types of data may be analysed with PLSR, either together with any of the four data types described from Section~\ref{sec:data_consumerLiking} through \ref{sec:data_sensory} or separately.\\

\subsection[Principal component regression]{Principal component regression}\label{sec:statMethod_PCR}
PCR~\citep{martens88} is another multivariate regression method that is well established in the field of sensometrics. PCR is basically a two-step procedure. First, PCA is applied to the $X$ matrix, finding principal components that explain the variance in the $X$ data only. Second, linear regression is applied to project the variables of $Y$ onto the PCA subspace of $X$. In this way the resulting components are influenced by the variation in $X$ only, unlike PLSR (see Section~\ref{sec:statMethod_PLSR}) where both $X$ and $Y$ influence the computation of components. \\
PCR as implemented in ConsumerCheck is coded in \textbf{Python} and provides $X$ scores, $X$ loadings, $Y$ loadings, $X$ \& $Y$ correlation loadings, calibrated and validated explained variances for for $X$ and $Y$ as well as predictions of $Y$ after a number of components for calibration and validation. If needed, further computation results, such as root mean square error of calibration (RMSEP) and cross validation (RMSECV), etc., are available from the PLSR class when using the \textbf{Python} source code directly outside ConsumerCheck. \\
As with PLSR above, PCR is accessible through the \textit{Prefmap} tab (see Figure~\ref{fig:GUI_prefmap} and details on the usage in Section~\ref{sec:GUI_PrefMap}) and the \textit{PLSR/PCR} tab (see Figure~\ref{fig:GUI_PLSR-PCR} and details on the usage in Section~\ref{sec:GUI_PLSR_PCR}). Under the \textit{Prefmap} tab the use of PCR is restricted to only \textit{consumer liking} data and \textit{descriptive analysis / sensory profiling} data, since \textit{preference mapping} deals only with these two types of data. Under the \textit{PLSR/PCR} tab other types of data may be analysed as for example \textit{consumer characteristics} data together with transposed \textit{consumer liking} data or \textit{product design} data together with \textit{consumer liking} data. If available, other types of data may be analysed with PLSR, either together with any of the four data types described from Section~\ref{sec:data_consumerLiking} through \ref{sec:data_sensory} or separately. \\

\subsection[Conjoint analysis]{Conjoint analysis}\label{sec:statMethod_Conjoint}
Conjoint analysis \citep{green71,green78} is a method for analysing the effects of \textit{design factors} (which are stored in the \textit{product design} matrix; see Section~\ref{sec:data_design}) and \textit{consumer characteristics} (see Section~\ref{sec:data_consumerCharacteristics}) on \textit{consumer likings} (see Section~\ref{sec:data_consumerLiking}). A common approach is to analyse it in a mixed effects model framework, where random effects consist of consumer effect and interactions between consumer effects and design factors, and fixed effects consist of design factors and consumer characteristics and possibly interactions between them. \\
Hence, in this type of analysis the following data set types are used: \textit{product design} matrices, \textit{consumer liking} matrices as well as \textit{consumer characteristics} matrices. Mixed effects models in conjoint analysis in ConsumerCheck are constructed using the \textbf{R} package \textbf{lme4} \citep{lme4}. The tests and post-hoc analysis for the models are performed using the \textbf{lmerTest} \textbf{R} package \citep{lmerTest}.  Conjoint Analysis as implemented in ConsumerCheck has a number of nice features: it can handle unbalanced data, multiple crossed effects, it can automatically find parsimonious models and perform post-hoc analysis. Different degrees of complexity (structure 1, 2 and 3) can be chosen by the user. More practical details on how to compute results and generate conjoint related plots in ConsumerCheck and how to interpret results are given in Section~\ref{sec:GUI_Conjoint}.

%

\subsection{Individual differences}\label{sec:statIndividDiff}
This tab is developed for analysing individual differences in consumer data in more detail with special emphasis on relating preference data to consumer attributes using PLS regression methods (see Section~\ref{sec:statMethod_PLSR}). All analyses under this tab use a PCA (see Section~\ref{sec:statMethod_PCA}) of the consumer liking data as point of departure regardless of whether it was meant for conjoint analysis or preference mapping.

There are three main analyses focused under the tab named Individual differences: The first is to relate loadings from the PCA of the liking data or the liking data themselves to consumer attributes (used as independent/input variables). This provides typical PLS plots showing relations between liking patterns and consumer attributes. The second analysis implemented focuses on a posteriori visual segmentation based on the PCA loadings \citep{naes18}. The segments are identified by encircling consumers in the plot using the cursor.  The segments identified in this way are then related to consumer attributes for interpretation using PLS-DA analysis (see Section~\ref{sec:statMethod_PLSR}). Again one obtains PLS based plots for visual interpretation of segments. The third option is visual interpretation of a priori defined segments marked in the consumer attributes data set (see Section~\ref{sec:data_consumerCharacteristics}). The different segments will show up in different colors in the plot.

\subsection[Future implementations of statistical methods]{Future implementations of statistical methods}\label{sec:statMethod_future}
ConsumerCheck is an ongoing project and there are plans to extend the ConsumerCheck software with more statistical methods.

\section[Software architecture]{Software architecture}\label{sec:SA}
The application framework and main part of the ConsumerCheck software is programmed in \textbf{Python}. The implemented statistical methods PCA, PLSR and PCR are adapted from the \textbf{Python} package \textbf{hoggorm} \citep{tomic19}, which was coded using the \textbf{Python} package \textbf{numpy} \citep{oliphant06}. Cross validation for these methods was carried out by use of the cross validation module from the \textbf{Python} package \textbf{scikit-learn} \citep{pedregosa11}. The \textbf{ETS} package (Enthought Tool Suite) was used for building the GUI and the functionality for handling user input and using statistical methods for analysis. Conjoint analysis in ConsumerCheck is implemented with the \textbf{R} package named \textbf{lmerTest} which is accessed by the framework through the \textbf{Python} based \textbf{PypeR} package \citep{Xia10} that interfaces \textbf{Python} and \textbf{R}.

\section[How to use ConsumerCheck]{How to use ConsumerCheck}\label{sec:GUI}
This section presents the graphical user interface to the reader and discusses all possible settings for data import, data summary and statistical analysis. A brief summary on the statistical methods are provided from Section~\ref{sec:statMethod_Basic} through ~\ref{sec:statMethod_Conjoint}. \\

\begin{figure}
  \centering
  \includegraphics[scale=0.6]{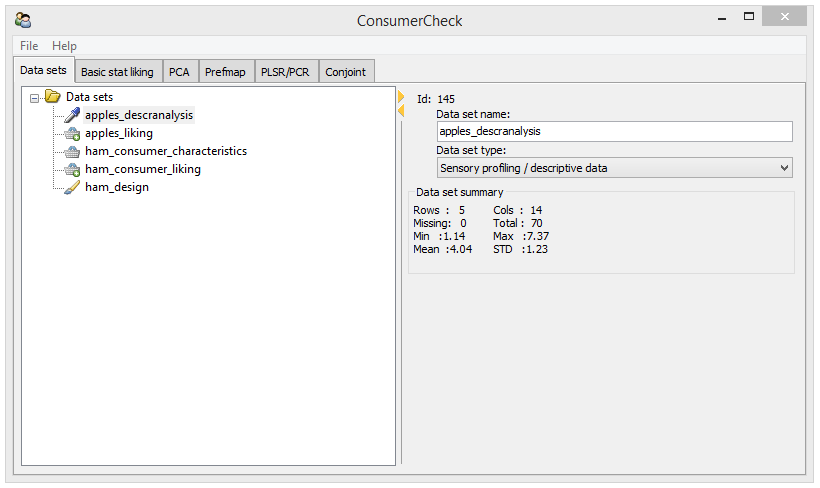}\\
  \caption{This is a screenshot of the graphical user interface of ConsumerCheck. The screenshot shows five data sets that were previously imported in the last ConsumerCheck session. These are the real world data that were described in Section~\ref{sec:data_used}}\label{fig:GUI_main}
\end{figure}

The main widget of the graphical user interface (GUI) is shown in Figure~\ref{fig:GUI_main}. In order to navigate from one statistical method to another there are a number of tabs at the top of the widget where each tab represents a statistical method except for the first one. The tabs are named: \textit{Data sets, Basic stat liking, PCA, Prefmap} and \textit{Conjoint}. All tabs have the same structure: (I) a so-called tree-control on the left side where the user can generate plots or tables by double clicking on an tree-control item; (II) a panel on the right side where various method-specific parameters can be set. In the following sub sections each tab will be explained in detail. First, however, following the chronological order of data analysis, the data need to be imported.

\subsection[Data import]{Data import and data removal}\label{sec:GUI_DataImport}
\subsubsection[Accepted file formats]{Accepted file formats}\label{sec:GUI_DataImport_fileFormats}
ConsumerCheck accepts several file formats for data import.
\begin{itemize}
 \item plain files such as \textbf{.txt} or \textbf{.csv}
 \item Excel files, both \textbf{.xls} and \textbf{.xlsx}
\end{itemize}

Note that ConsumerCheck remembers which data sets were imported in the last session and automatically loads them when ConsumerCheck is launched. When launching ConsumerCheck for the very first time no data are imported. One can import data by selecting \textit{File -> Add Data sets} from the menu at the top of the GUI. When importing data for the first time after launching ConsumerCheck, regardless of whether data are already imported or not, a window appears providing short information on how each type of data should be structured. The information provided in this window is a short summary of what is described in Section~\ref{sec:dataType}. Then a standard \textit{Open file} dialog appears which allows for selection of one or more files for import. After clicking the \textit{Open file} button in the open file dialog an import dialog appears for each selected file in successive order. The look and type of the import dialog depends on the format of the selected file and as such provides different parameter settings for the import. Below a short description of the import settings for text and Excel files is given.

\begin{figure}
  \centering
  \includegraphics[scale=0.6]{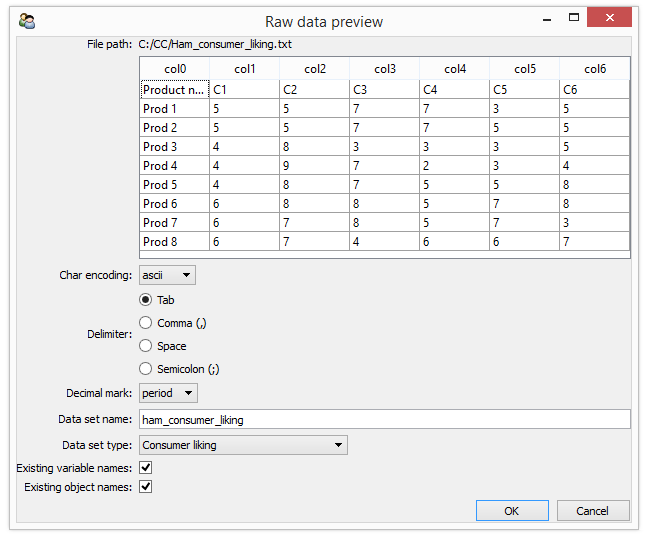}\\
  \caption{This is a screenshot of the import dialog for data stored in plain files such as \textbf{.txt} or \textbf{.csv}.}\label{fig:GUI_import_text}
\end{figure}

\subsubsection[Text files]{Data import dialog for text files}\label{sec:GUI_DataImport_textFilesGUI}
Figure~\ref{fig:GUI_import_text} shows a screenshot of the data import dialog for text files. At the top of the widget the path to the location of the file is displayed. Below is a grid providing a preview of the raw data that is about to be imported. This may be useful for quick inspection of whether the correct data were selected for import. The next import parameter provides a drop down menu where the encoding of the data may be selected. By default \emph{ASCII} encoding is selected which fine to use if the data files do not contain special characters. The other two choices are \textit{UTF-8} and \textit{latin-1}. For more information on which encoding should be used, please consult (http://www.unicode.org/). Below there are three so-called radio buttons where the user can communicate to ConsumerCheck how each column of data in the text file is separated from one another, that is by 'tab', 'comma' or 'space'. Next, the user can set whether floats (that is numbers with decimals) are defined by commas or periods. Below, users can provide a name for the data set that will be used throughout ConsumerCheck. If the user doesn't give the data set a new name at this point, it is still possible to do so in the \textit{Datas set} tab, that is described in Section~\ref{sec:GUI_Datasets}. Beneath there is a drop down menu where the user can define of which type the data set is. The five possible selections in the drop down menu are \textit{consumer liking}, \textit{consumer characteristics}, \textit{product design}, \textit{descriptive analysis / sensory profiling} and \textit{other}. Those are the data types discussed from Section~\ref{sec:data_consumerLiking} through ~\ref{sec:data_sensory} and ~\ref{sec:data_other}. Selection of data type is important in order to have ConsumerCheck recognise appropriate data that are suitable for a chosen statistical method. If the user does not set the data type in the import dialog it is still possible to do so later in the \textit{Data sets} tab (see section ~\ref{sec:GUI_Datasets}). Eventually, the user can check or uncheck two check boxes to indicate whether the data have product and variable names included or not.

\begin{figure}
  \centering
  \includegraphics[scale=0.6]{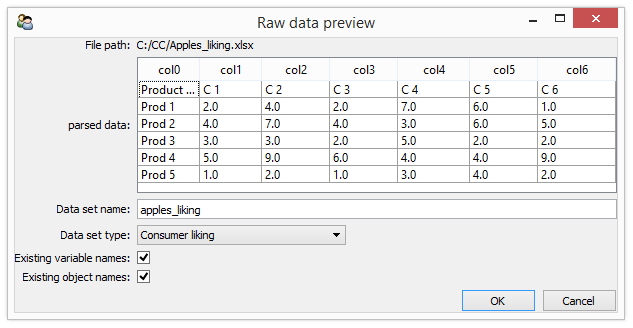}\\
  \caption{This is a screenshot that shows the import dialog for data stored in Excel files.}\label{fig:GUI_import_Excel}
\end{figure}

\subsubsection[Excel files]{Data import dialog for Excel files}\label{sec:GUI_DataImport_ExcelFilesGUI}
Figure~\ref{fig:GUI_import_Excel} shows the dialog for data import from Excel files. Its structure and usage are almost identical to the import dialog for text files, except for that there are no settings for text encoding and options for delimiter. Both are detected automatically by ConsumerCheck.

\subsubsection[Missing values]{Missing values in data}\label{sec:GUI_DataImport_missingValues}
\begin{figure}
  \centering
  \includegraphics[scale=0.6]{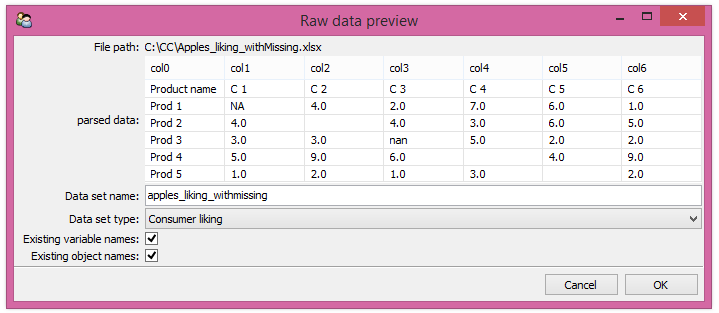}\\
  \caption{This is a screenshot of the Excel import dialog of data with missing values. Note that there are missing values in the following positions of the shown data: col1-row2 marked as \textbf{NA}; col2-row3, col4-row5 and col5-row6 as \textit{empty cells}; col3-row4 marked as \textbf{nan}. All three are valid approaches to mark missing values in the data.}\label{fig:GUI_missing_prior}
\end{figure}

As mentioned earlier (see Section~\ref{sec:dataType}) the import of data with missing values is allowed. There are plans to implement imputation routines for the handling missing values in future versions of ConsumerCheck, but progress will greatly depend on availability of funding and resources in general. Until then, users need to impute missing values with their preferred imputation method outside ConsumerCheck before importing the data into the software. When data with missing values are to be imported, the missing values may be indicated either by leaving their respective cells empty or by marking them as \textbf{NA} or \textbf{nan}. Figure~\ref{fig:GUI_missing_prior} shows an example of importing data with missing values. Although none of the implemented statistical methods can handle data with missing values yet, providing the possibility to do so is a first step in that direction and will be built upon in future versions of ConsumerCheck.

\subsubsection[Additional visualisation information]{Adding information for richer visualisation in plots}\label{sec:GUI_DataImport_addInfo}
\begin{figure}
  \centering
  \includegraphics[scale=0.6]{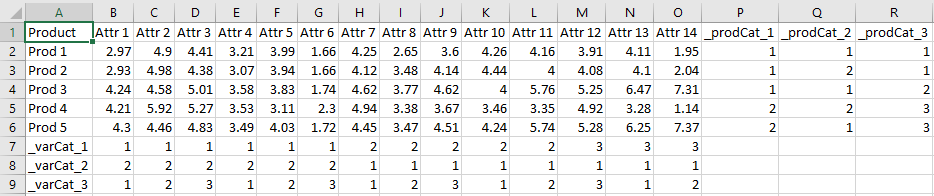}\\
  \caption{Apples descriptive analysis data with additional information on samples and variables shown in an Excel sheet before importing it into ConsumerCheck. In this example three columns were added that represent three specific product categories, as well as three extra rows that represent three specific variable categories. Note that the names of such additional columns must start with an underscore for ConsumerCheck to recognise it as such.}\label{fig:GUI_additional_info}
\end{figure}

ConsumerCheck offers an option for implementing additional information that colours scores and loadings of PCA results (Section~\ref{sec:statMethod_PCA}) according to user-defined groups of samples and variables. Colouring scores and loadings based on such user-defined groups of samples or user-defined groups of variables can make interpretation of the results a lot easier. For samples scores such user-defined groups could be based on for example the product information from experimental designs and are made known to ConsumerCheck by adding columns to the original data as shown in Figure~\ref{fig:GUI_additional_info}. Note that such additional columns holding such user-defined information must start with an underscore. In a similar way, for variable loadings user-defined groups could be based on variable type, as for example whether an attribute represents an odour or flavour and make them known to ConsumerCheck by adding rows indicating which group each variable belongs to. Note that also here the additional rows holding user-defined group information also must start with a underscore.

In Section~\ref{sec:GUI_PCA_scores} and \ref{sec:GUI_PCA_loadings} examples will show how this additional information can be utilised using drop down menus in the plots resulting in coloured PCA scores and loadings plots.

\subsubsection[Removing data]{Removing data}\label{sec:GUI_DataImport_dataRemove}
If a previously imported data set needs to be removed from ConsumerCheck this can be done easily by selecting the following from the main menu: \textit{File -> Remove Data sets}.

\subsection[Data sets tab]{Data sets tab}\label{sec:GUI_Datasets}
The \textit{Data sets} tab is the first of several tabs of the GUI. Its main purpose is to provide a general overview of the data imported, a short summary of each data, tools for setting data parameters and methods for processing data. The \emph{Data sets} tab, like all other tabs in the GUI, is divided into a left and right panel. Figure~\ref{fig:GUI_main} shows an example of what the the \emph{Data sets} tab looks like with five data set imported. These are the apple and ham data that were described earlier in Section~\ref{sec:data_appleData} and \ref{sec:data_hamData}, respectively.

The left panel shows a so-called tree-control with one data set at each branch. With a single left-click on a data set information that is specific for this data set is shown in the right panel, i.e. the \textit{data set name}, \textit{data set type} and \textit{data set summary}. Both the name and the type of the data may have been set already in their respective import dialogs (see Section~\ref{sec:GUI_DataImport}), but here the user can change these parameters again if needed. In the \textit{data set name} text field at the top of the right panel the name of the particular data set may be changed. The name defined in this text field is then used in the statistical method specific tabs in ConsumerCheck (see Sections~\ref{sec:GUI_Basic} to~\ref{sec:GUI_Conjoint}). With the \textit{data set type} drop-down menu right below the type of the data set may be set. Below, the \textit{data set summary} provides a short summary of the respective data, such as the dimension of the data, the mean and standard deviation across all entries as well as the minimum and maximum values in the data. A double left-click on a data set in the tree-control generates a new window that visualises the data in a sheet. From that window one can copy the data by clicking on the \textit{copy to clipboard} button and paste it into other software applications such as Excel or Open Office Calc Spreadsheet. A single right-click on a data set invokes a menu that lets the user do various things with the data. At the time of writing, this menu contains two options, but more may follow in future versions of ConsumerCheck: (I)~\textit{Create transposed copy} and (II)~\textit{Delete}. The first option allows the user to make a transposed copy of the selected data set, meaning that a copy of that specific data is added at the lower end of the tree-control, but where rows have become columns and columns have become rows. This may be useful when applying the multivariate statistical regression methods \textit{PLSR} (see Section~\ref{sec:statMethod_PLSR}) and \textit{PCR} (see Section~\ref{sec:statMethod_PCR}) to two data sets as it is done with the \textit{PLSR/PCR} tab (see Section~\ref{sec:GUI_PLSR_PCR}). The second option lets the user delete data sets from ConsumerCheck. The data then are no longer available at the tree-control. If needed again, the data may be re-imported the usual way as described in Section~\ref{sec:GUI_DataImport}.

\subsection[Basic stat liking tab]{Basic statistics for consumer liking data}\label{sec:GUI_Basic}
\begin{figure}
  \centering
  \includegraphics[scale=0.6]{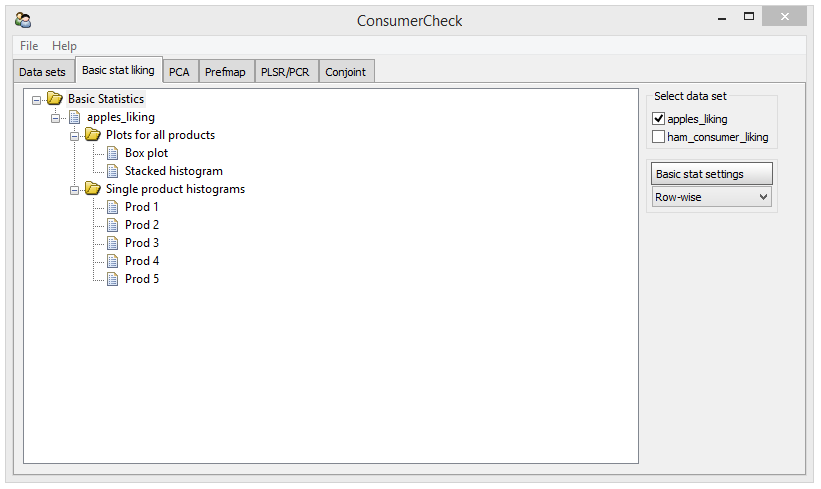}\\
  \caption{This screenshot shows the graphical user interface of the \textit{Basic stat liking} tab.}\label{fig:GUI_basicStatLiking}
\end{figure}

The purpose of \emph{Basic stat liking} tab (see screenshot in Figure~\ref{fig:GUI_basicStatLiking}) is to provide visualisation and simple analysis of consumer liking data to the user. This implies that only data of type \textit{consumer liking} are listed in the \textit{Select data set} box in the upper right corner of the GUI and as such are available for visualisation and analysis. As seen in the Figure~\ref{fig:GUI_basicStatLiking}, in this case consumer liking data from the apple and ham data set are present and available for visualisation. At the left there is a tree-control from which plots may be generated by double left-clicking on tree-control items. The tree-control is dynamic and expands or retracts as consumer liking data are checked or unchecked. The tree-ctrl provides three types of plots: \textit{box plots}, \textit{stacked histogram plots} and \textit{single product histogram plots}. Each type will be described below.

\subsubsection[Box plot]{Plots for all products -  Box plot}\label{sec:GUI_Basic_boxPlot}
\begin{figure}
  \centering
  \includegraphics[scale=0.6]{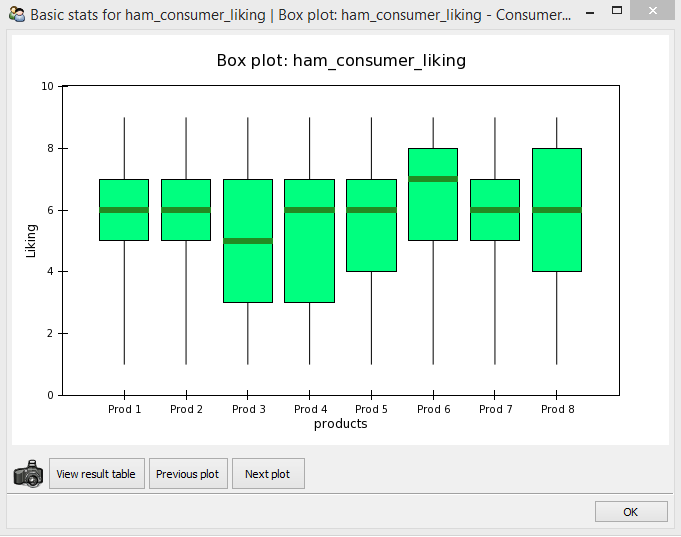}\\
  \caption{This is an example of what the box plot looks like for the \textit{ham consumer liking} data. The plot is found under the tab named \textit{Basic stats liking}}\label{fig:GUI_basicStatLiking_boxplot}
\end{figure}

Figure~\ref{fig:GUI_basicStatLiking_boxplot} shows the box plot for the \textit{ham consumer liking} data where consumers rated 8 food products on a hedonic scale from 1 to 9, where 1 represents "don't like at all" and 9 represent "like very much". The box plot shows that across all consumers each product received the highest (9) and lowest (1) rate by at least one consumer. This is visualised by the vertical lines that extend from 1 to 9 for each product. The green boxes for each line visualise the distribution of the ratings between the 25th and 75th percentile. The dark green line across the green boxes shows the median rating for that product. Note that the plot can be saved in .png format when clicking on the photo camera icon placed in the lower left corner of the plot window. This is a common feature for all plots implemented in ConsumerCheck. Another common feature is the \textit{View result table} button to the right of the photo camera icon. By clicking on it, a new window appears showing a data sheet that holds all the numbers a plot is based on. In case of the box plot, the data sheet contains \textit{max}, \textit{min}, \textit{median}, \textit{25th} and \textit{75th percentile} of the ratings of every tested product

\subsubsection[Stacked histogram]{Plots for all products - stacked histograms}\label{sec:GUI_Basic_stackedHisto}
\begin{figure}
  \centering
  \includegraphics[scale=0.6]{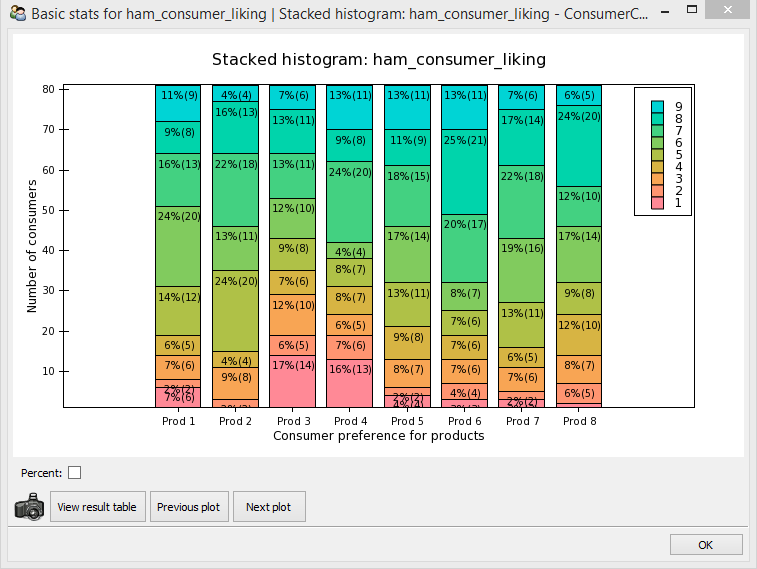}\\
  \caption{This is an example of what a stacked histogram \textit{ham consumer liking} data. The plot is found under the \textit{Basic stat liking} tab.}\label{fig:GUI_basicStatLiking_stackedHistogram}
\end{figure}

Stacked histograms provide another and richer way of visualising \textit{consumer liking} data. In Figure~\ref{fig:GUI_basicStatLiking_stackedHistogram} a stacked histogram plot is shown for the same data as presented earlier in a box plot in Figure~\ref{fig:GUI_basicStatLiking_boxplot}. Along the horizontal axis again the products are shown, while the vertical axis displays either the number of consumers or a percentage of the total number of consumers. If percentages are to be shown, the \textit{Percent} checkbox at the bottom left corner of the window needs to be checked. With the stacked histogram each bar represents one product and each colour in the bar represents a certain rating of the product. For \emph{prod 3} one can see that 14 consumers or 17\% of the total number of consumers rated this product with 1 ("don't like at all"). 5 consumers or 6\% of the consumers rated \emph{prod 3} with 2, and so on. In this way the distribution of the ratings is visualised in a more detailed way than in the box plots.

\subsubsection[Histograms]{Single product histograms}\label{sec:GUI_Basic_basicHisto}
\begin{figure}
  \centering
  \includegraphics[scale=0.6]{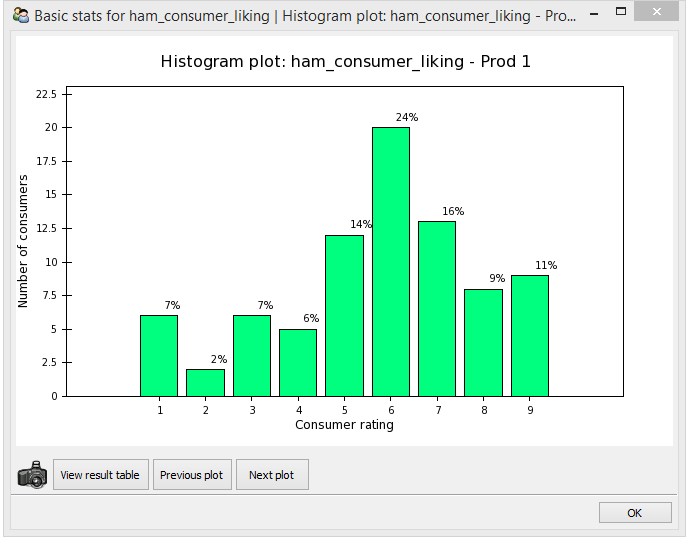}\\
  \caption{This screenshot shows the histogram for product \textit{prod 1} of the \textit{ham consumer liking} data.}\label{fig:GUI_basicStatLiking_histogram}
\end{figure}

Single product histograms show for each product the distribution of the liking ratings in separate histograms. Figure~\ref{fig:GUI_basicStatLiking_histogram} shows an example for \emph{prod 1}.

Now, instead of putting all information into one bar as seen in the stacked histogram plot, one plot is dedicated to \emph{prod 1} alone. In the single product histogram the bars represent increasing liking rates from left to right. For each rating the percentage of consumers having rated the product this way is displayed on the top of the bar.

\subsubsection[Column-wise]{Column-wise summary of consumer liking data}\label{sec:GUI_Basic_colWise}
The main idea behind the \textit{Basic stat liking} tab is to provide information on the distribution of liking ratings with focus on the products / objects in the \textit{consumer liking} data. It is, however, possible to visualise the liking distributions also with focus on the consumer. This can be achieved by selecting \textit{Column-wise} in the \textit{Basic stat settings} drop-down menu at the right side of the GUI. Double clicking on \textit{Box plot} and \textit{stacked histogram} in the tree-control now generates box plots and stacked histograms, respectively, for each consumer. Note that the default for the \textit{Basic stat setting} is \textit{row-wise}, i.e. focus on the products.

\subsection[PCA tab]{Principal component analysis}\label{sec:GUI_PCA}
\begin{figure}
  \centering
  \includegraphics[scale=0.6]{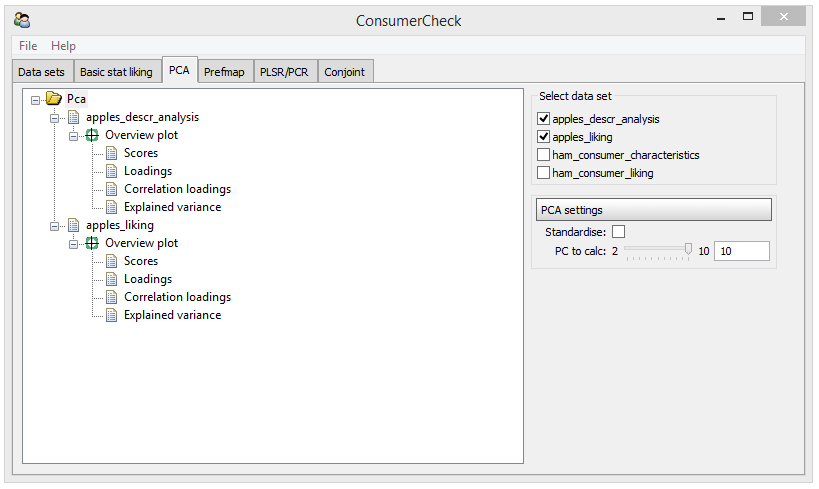}\\
  \caption{A screenshot of the \textit{PCA} tab.}\label{fig:GUI_PCA}
\end{figure}

When selecting the PCA tab the user can carry out principal component analysis on data of type \textit{descriptive analysis / sensory profiling}, \textit{consumer liking} and \textit{consumer characteristics}. Figure~\ref{fig:GUI_PCA} shows an example of the PCA tab with four of data sets ready for analysis. Throughout the PCA section the \textit{apple descriptive analysis / sensory profiling} data (as described in Section~\ref{sec:data_used}) will be used to illustrate what results are provided when analysing data with PCA. Again, the tree-control for generating plots is on the left side and the data available for analysis with PCA are listed in the upper right corner under \textit{Select data set}. Below, the user can choose more settings for the computation of the PCA model. By checking the \textit{Standardise} checkbox, all variables in the data are standardised such that they have zero mean and a standard deviation that equals one. By default, that is when the \textit{Standardise} checkbox is unchecked, variables are mean centered. Note that variables with zero variance across objects/rows are left out of analysis when the \textit{Standardise} checkbox is checked. This is because variables with zero variance cannot be standardised (division by zero). In such a case a message box will inform the user about leaving out such a variable. Moreover, the user can select how many principal components (PC's) are to be computed for the PCA model. This can be done either by typing the number of wanted PC's directly into the text box or by using the slider at its left side. Note that the maximum number of PC's that can be computed from a data set is equal to either \textit{number of variables} or \textit{number of objects} in the data set, whichever is smaller. Below, we will discuss which plots may be generated from the tree control.

\subsubsection[PCA overview plot]{PCA - Overview plot}\label{sec:GUI_PCA_overview}
\begin{figure}
  \centering
  \includegraphics[scale=0.3]{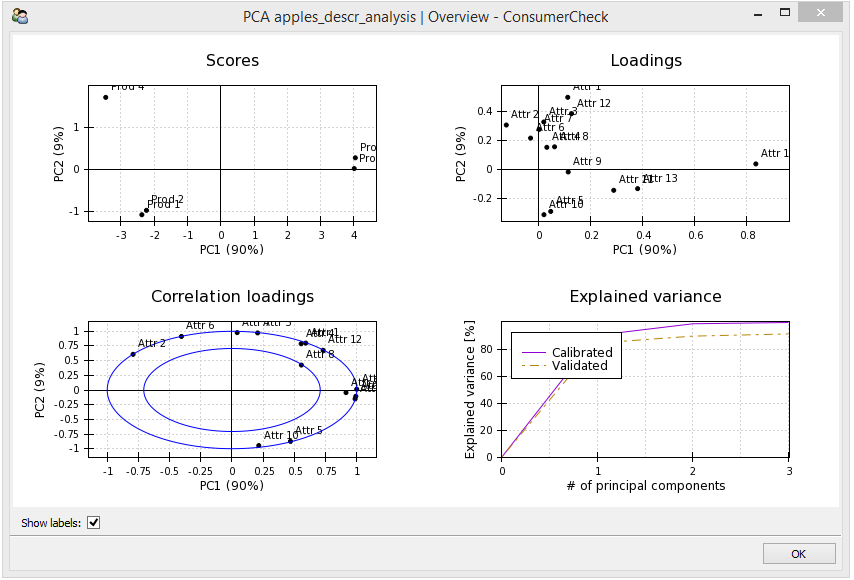}\\
  \caption{Screenshot of an \textit{Overview plot} for the PCA model based on the \textit{apple descriptive analysis / sensory profiling} data.}\label{fig:GUI_PCA_overview}
\end{figure}

By double-clicking on the tree control item named \textit{PCA overview plot} a new window will appear that consists of four sub-plots. An example of such an overview plot is shown in Figure~\ref{fig:GUI_PCA_overview}. As can be seen the four sub-plots are PCA \textit{Scores}, \textit{Loadings}, \textit{Correlation loadings} and \textit{Explained variance}. More details on each plot are given below in the respective subsections. With a single left-click on one of the four subplots a new window will appear showing an enlarged version of that specific plot. The same can be achieved by directly double left-clicking on the respective item in the tree control.

\subsubsection[PCA scores]{PCA scores}\label{sec:GUI_PCA_scores}
\begin{figure}
  \centering
  \includegraphics[scale=0.6]{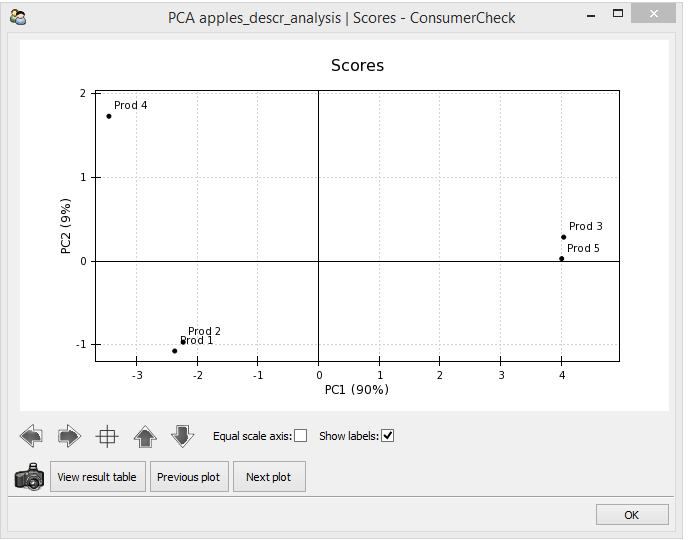}\\
  \caption{Screenshot of a PCA scores plot showing how the five products from the \textit{apple descriptive analysis / sensory profiling} data are distributed across PC1 and PC2. PC1 and PC2 together describe 99\% of the variation in the data.}\label{fig:GUI_PCA_scores}
\end{figure}

The PCA scores plot visualises how the objects or products from the analysed data matrix are distributed across the space spanned by two principal components (PC). By default the plot shows the scores for PC1 and PC2, that is the components explaining the highest and next to highest variance in the data. Figure~\ref{fig:GUI_PCA_scores} shows a example of what a PCA scores plot may look like. Here PC1 and PC2 explain 90\% and 9\% of the calibrated variance in the data, respectively, totalling 99\%. In other words, almost all of systematic variation in the data is visualised by these two components. From the plot we can see that product 3 and 5 are very similar since they are located very close to each other. At the other side of the plot there are products 1, 2 and 4 indicating that these products are very different from product 3 and 5 given the fact that PC1 explains 90\% of the calibrated explained variance. Product 1 and 2 are very similar because of their proximity in the plot. Product 4 is not too different from product 1 and 2 with regard to PC1, but some differences are present since the products are spread out across PC2 which accounts for 9\% of the variance in the data.\\

\begin{figure}
  \centering
  \includegraphics[scale=0.8]{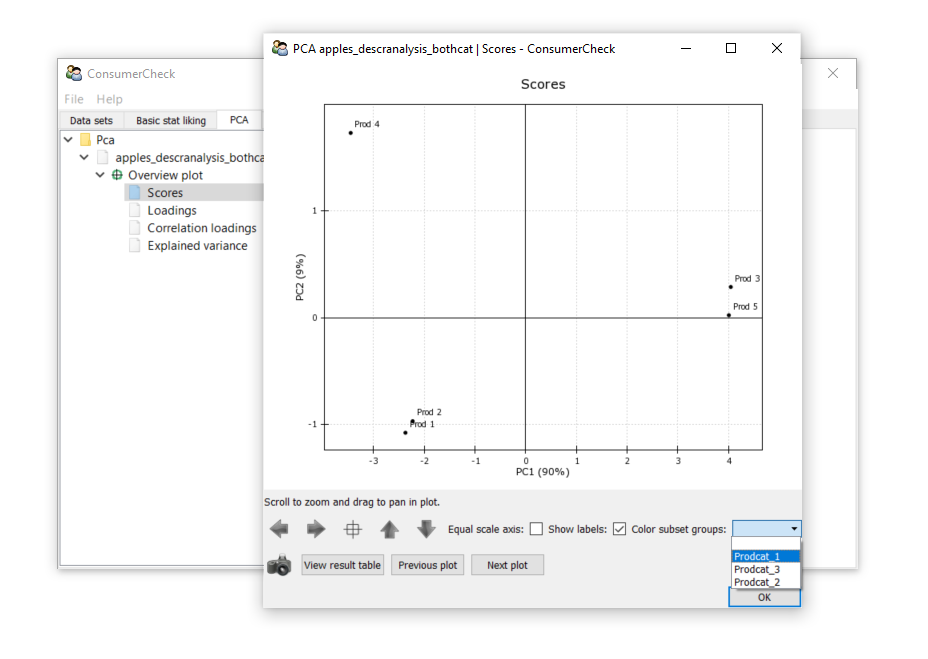}\\
  \caption{Using the drop-down menu, the user can select by which additional information the scores should be coloured.}\label{fig:GUI_PCA_scores_dropDown}
\end{figure}

\begin{figure}
  \centering
  \includegraphics[scale=0.6]{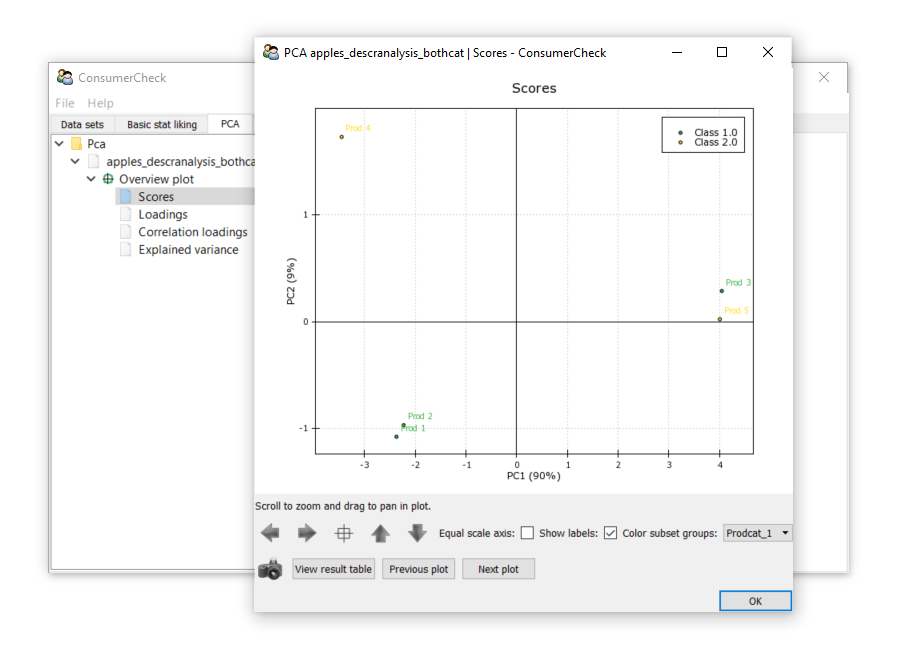}\\
  \caption{The scores are now coloured according to the selected user-defined group.}\label{fig:GUI_PCA_scores_coloured}
\end{figure}

Figure~\ref{fig:GUI_PCA_scores_dropDown} and \ref{fig:GUI_PCA_scores_coloured} illustrate how to colour the PCA scores by user-defined information. How user-defined information is added to the data to allow for such colouring is described in detail in Section~\ref{sec:GUI_DataImport_addInfo}. Using the drop-down menu in the lower right corner of the window, the user can select by which user-defined group the scores shall be coloured.

\subsubsection[PCA loadings]{PCA loadings}\label{sec:GUI_PCA_loadings}
\begin{figure}
  \centering
  \includegraphics[scale=0.8]{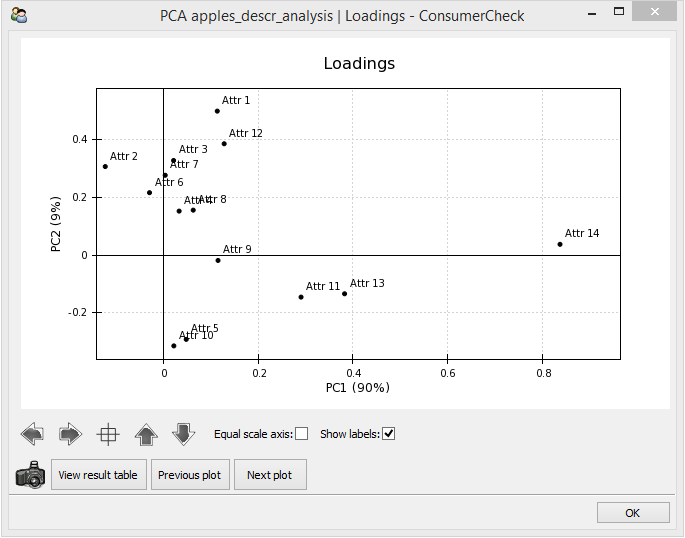}\\
  \caption{Screenshot of a PCA loadings plot for the \textit{apple descriptive analysis / sensory profiling} data.}\label{fig:GUI_PCA_loadings}
\end{figure}

The PCA loadings plot visualises how the variables, which in our case are sensory attributes describing the food product, contribute to the variation in the data. Figure~\ref{fig:GUI_PCA_loadings} shows the PCA loadings for the \textit{apple descriptive analysis / sensory profiling} data. Clearly, attribute 14 contributes much to the variation explained by PC1 since it has a large absolute loading for PC1 compared to all other sensory attributes. Attribute 11 and 13 are two other variables contributing much to the variation explained by PC1. For PC2, attributes 5 and 10 on one side and attributes 1, 2, 3, 7 and 12 on the opposite side are variables contributing most to variation. In general, variables that are located close to each other are highly correlated to one another with respect to the plotted principal components and vice versa. The closer a variable to the origin, the less it contributes to systematic variation explained by the two visualised PC's.

By superimposing the PCA scores and loadings plot one can get more information on the products. Both attribute 14 and products 3 and 5 are located on the right side of the loadings and scores plot, respectively. This means that these two products have high values for attribute 14 while products 1, 2 and 4 have lower values for attribute 14, since they are located on the opposite side with regard to PC1. Note that in this example the data are not standardised, which is why attribute 14 is dominating. The reason for not standardising the variables here is that the trained sensory panel scores all attributes on the same scale, which in our case is from 1 (low intensity) to 9 (high intensity). Elaborating this data further we can see that product 1 and 2 have high intensities for attribute 5 and 10 and lower intensities of attribute 1 and 12 (which are located at the opposite side with regard to PC2). For product 4 the opposite is true. \\

\begin{figure}
  \centering
  \includegraphics[scale=0.8]{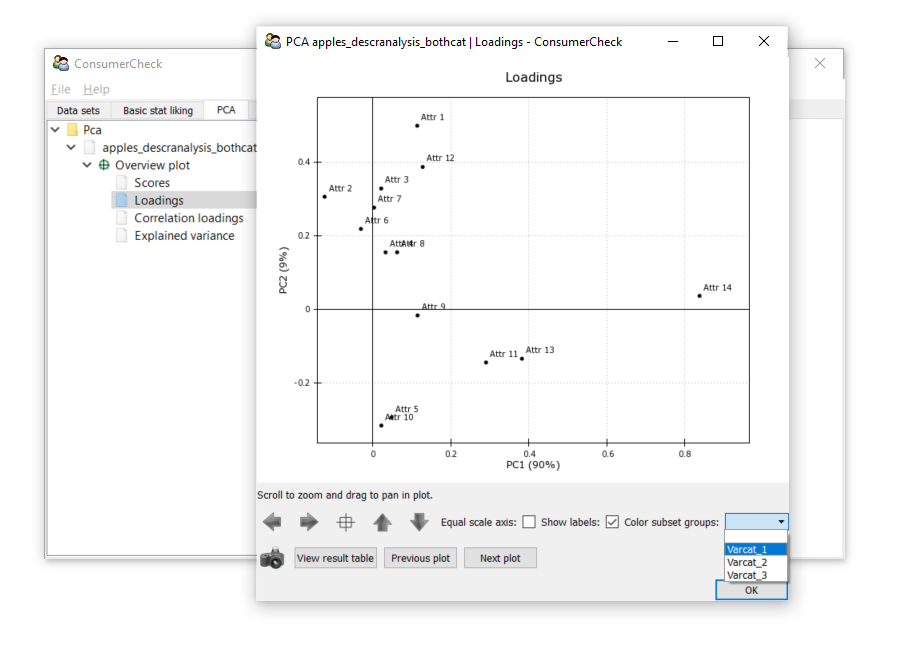}\\
  \caption{Using the drop-down menu, the user can select by which additional information the loadings should be coloured.}\label{fig:GUI_PCA_loadings_dropDown}
\end{figure}

\begin{figure}
  \centering
  \includegraphics[scale=0.8]{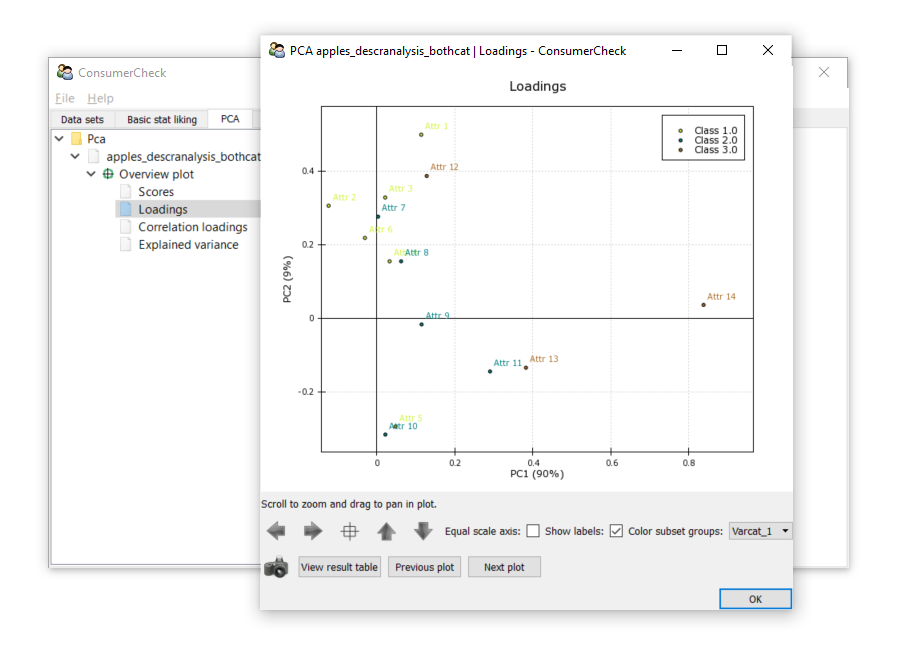}\\
  \caption{The loadings are now coloured according to the selected user-defined group.}\label{fig:GUI_PCA_loadings_coloured}
\end{figure}

Figure~\ref{fig:GUI_PCA_loadings_dropDown} and \ref{fig:GUI_PCA_loadings_coloured} illustrate how to colour the PCA loadings by user-defined information. How user-defined information is added to the data to allow for such colouring is described in detail in Section~\ref{sec:GUI_DataImport_addInfo}. Using the drop-down menu in the lower right corner of the window, the user can select by which user-defined group the loadings shall be coloured.

\subsubsection[PCA correlation loadings]{PCA correlation loadings}\label{sec:GUI_PCA_corrLoadings}
\begin{figure}
  \centering
  \includegraphics[scale=0.6]{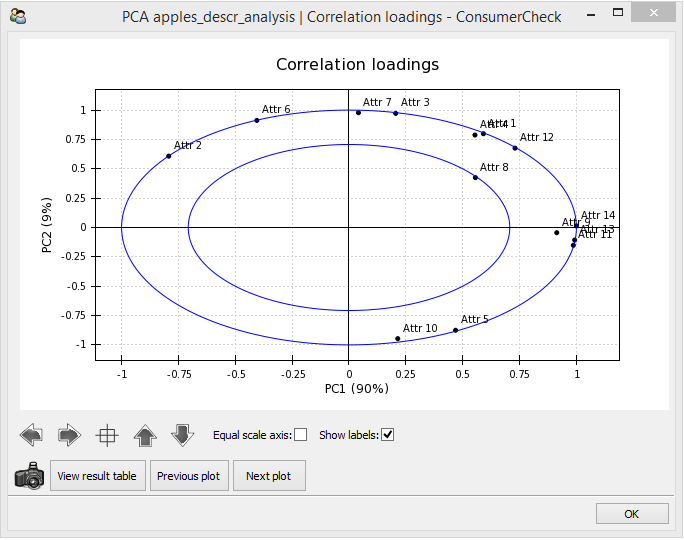}\\
  \caption{PCA correlation loadings for the \textit{apple descriptive analysis / sensory profiling} data.}\label{fig:GUI_PCA_corrLoadings}
\end{figure}

PCA correlation loadings, as shown in Figure~\ref{fig:GUI_PCA_corrLoadings}, are another way of visualising the contribution of the variables to the total variance in the data. PCA correlation loadings provide information on how systematic the variance of a variable is with regard to the computed PC's, not only how much variance was contributed by the variable (as visualised in the PCA scores plot). More precisely, a correlation loading is actually the correlation between the original data of a specific variable and the scores of a specific PC \citep{martens01,mardia79}. In this way one can see to which degree the variation from a specific variable is systematic or rather noisy, regardless of the total variance it contributes. The two rings in the correlation loadings plot in Figure~\ref{fig:GUI_PCA_corrLoadings} indicate specific amounts of explained variance for the attributes at hand. The outer ring represents 100\% explained variance while the inner ring represents 50\% explained variance.

Consider an example with attributes 4 and 8. When looking at the loadings plot in Figure~\ref{fig:GUI_PCA_loadings} these two variables are located close to each other contributing about the same amount of variance to the variance explained by PC1 and PC2. In the correlation loadings plot (Figure~\ref{fig:GUI_PCA_corrLoadings}), however, they are no longer located close to each other. Attribute 8 is located just inside the inner ring, which indicates that just under 50\% of the variation of this variable is explained by PC1 and PC2. Remember that that PC1 and PC2 together explain 99\% of the total variance in the data and that remaining higher PC's provide very little or no information. This means that not much more than 50\% of the variance of attribute 8 will be explained by the higher PC's and that the remaining variance of that variable is likely to be noise. Attribute 4 is very close to the outer ring, meaning that almost 100\% of its variation is explained by PC1 and PC2, thus indicating that its variance is very systematic. In this way it is possible to see that the variation of attribute 4 is much more systematic with the variance described by PC1 and PC2 than that of attribute 8, even though both of them contribute about the same amount of variance to the data. Considering this, PCA correlation loadings are a useful complement to the PCA loadings for better understanding of how variables contribute to the total variance in the data.

\subsubsection[PCA explained variances]{PCA explained variances}\label{sec:GUI_PCA_explVar}
\begin{figure}
  \centering
  \includegraphics[scale=0.6]{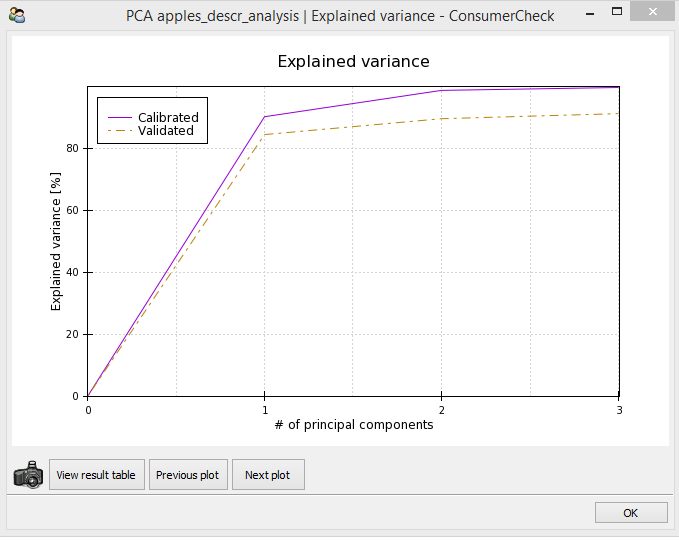}\\
  \caption{Cumulative calibrated and validated explained variance of the PCA model for the \textit{apple descriptive analysis / sensory profiling data}.}\label{fig:GUI_PCA_explainedVariance}
\end{figure}

Figure~\ref{fig:GUI_PCA_explainedVariance} shows the cumulative calibrated and validated explained variances of the same data. It can be seen easily that with only two PC's almost all of the variance in the data is explained by the model (99\% as mentioned above). Full cross-validation (also known as leave-one-out) was applied to the data for model validation. The resulting cumulative validated explained variance rises to about 85\% with PC1 and then slightly increases to 90\% after PC2, closely following the line for the cumulative calibrated explained variance and hence confirming that the model is robust.

\subsection[Prefmap tab]{Preference mapping}\label{sec:GUI_PrefMap}
Preference mapping is applied simultaneously to \textit{consumer liking} data and \textit{descriptive analysis / sensory profiling} data. The aim is to find drivers of liking that may determine why some products are preferred over other. The two standard statistical tools applied to build a preference mapping model are partial least squares regression (PLSR) and principal component regression (PCR). Both are implemented in ConsumerCheck and are coded in \textbf{Python}. \\

\begin{figure}
  \centering
  \includegraphics[scale=0.3]{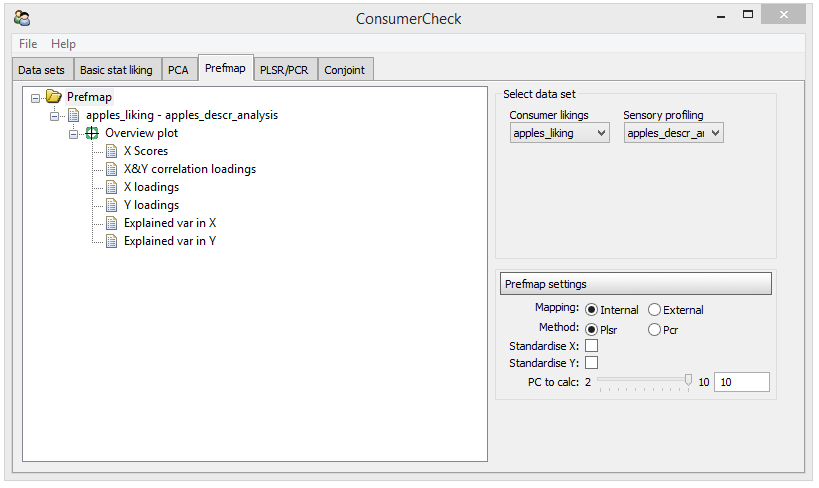}\\
  \caption{Screenshot of the GUI for \textit{preference mapping}. The tree control to the left is shows which plots may be generated for the preference mapping model based on the \textit{apple consumer liking} data and \textit{apple descriptive analysis / sensory profiling data}.}\label{fig:GUI_prefmap}
\end{figure}

When building a preference mapping model, both consumers and the trained sensory panel need to evaluate the same set of products. In each data, the row order of the products needs to be identical otherwise wrong data are linked together and results will lead to incorrect conclusions. Two drop down menus on the right side of the GUI let the user define which data are to be linked (see Figure~\ref{fig:GUI_prefmap}). The left drop down menu contains all data imported into ConsumerCheck that were tagged as \textit{consumer liking} data whereas the right drop down menu contains all data tagged as \textit{descriptive analysis / sensory profiling} data. If the number of rows in the two data set do not match because they originate from different experiments, ConsumerCheck will give an error message. Once two matching data set are selected a tree control appears on the left side of the GUI.
There are a few more settings for computation of the model at the right side of the GUI that can be adjusted by the user. First, the user may choose between internal preference mapping (\textit{consumer liking} data are set as the $X$ matrix in the model and the \textit{descriptive analysis / sensory profiling} data are set as the $Y$ matrix) or external preference mapping (\textit{descriptive analysis / sensory profiling} data are set as $X$ and \textit{consumer liking} data are set as Y). With the next setting the user may choose between the statistical methods PLSR or PCR. Below there are two check boxes where the user may choose to standardise either or both $X$ and Y. By checking the \textit{Standardise} checkboxes, all variables in the data are standardised such that they have zero mean and a standard deviation that equals one. By default, that is when the \textit{Standardise} checkboxes are unchecked, variables are mean centered. Note that variables with zero variance across objects/rows are left out of analysis when the respective \textit{Standardise} checkbox is checked. This is because variables with zero variance cannot be standardised (division by zero). In such a case a message box will inform the user about leaving out such a variable. With the last parameter setting the user can select the number of components to be computed for the model.

\subsubsection[Prefmap overview plot]{Preference mapping - overview plot}\label{sec:GUI_PrefMap_overview}
\begin{figure}
  \centering
  \includegraphics[scale=0.3]{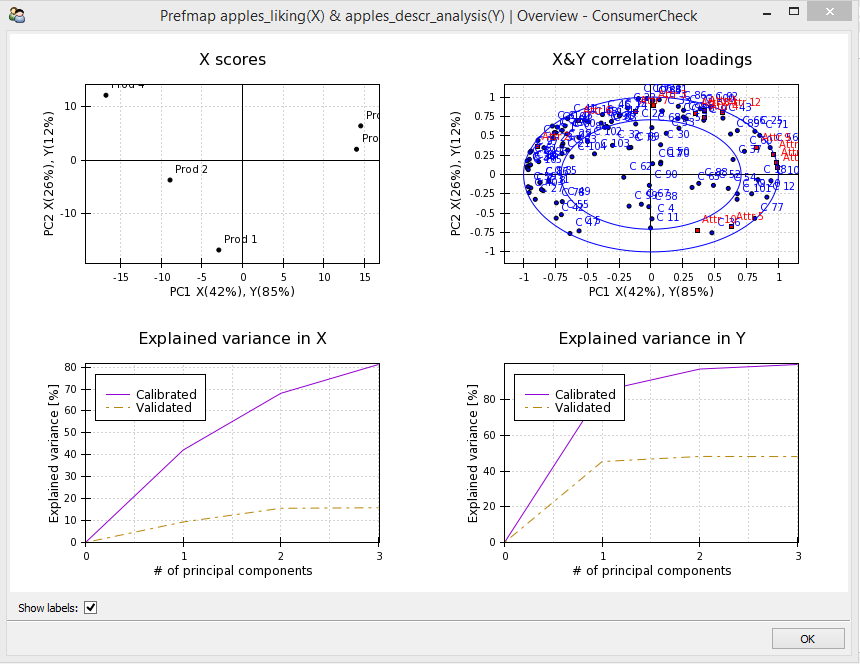}\\
  \caption{Overview plot for preference mapping model for the \textit{apple consumer liking} data and \textit{apple descriptive analysis / sensory profiling} data.}\label{fig:GUI_prefmap_overview}
\end{figure}

By left double-clicking on \textit{Overview plot} a new window opens that shows the $X$ scores (upper left), $X$ \& $Y$ correlation loadings (upper right), cumulative explained variance in $X$ (lower left) and cumulative explained variance in $Y$ (lower right) as shown in Figure~\ref{fig:GUI_prefmap_overview}. The respective plots can be accessed directly by left-double clicking on their respective item in the tree control or by a single left-click directly on the overview plot. More information on each plot is provided below. Furthermore, from the tree control separate plots for $X$ and $Y$ correlation loadings may be generated.

\subsubsection[Prefmap X scores]{Preference mapping - $X$ scores}\label{sec:GUI_PrefMap_Xscores}
\begin{figure}
  \centering
  \includegraphics[scale=0.6]{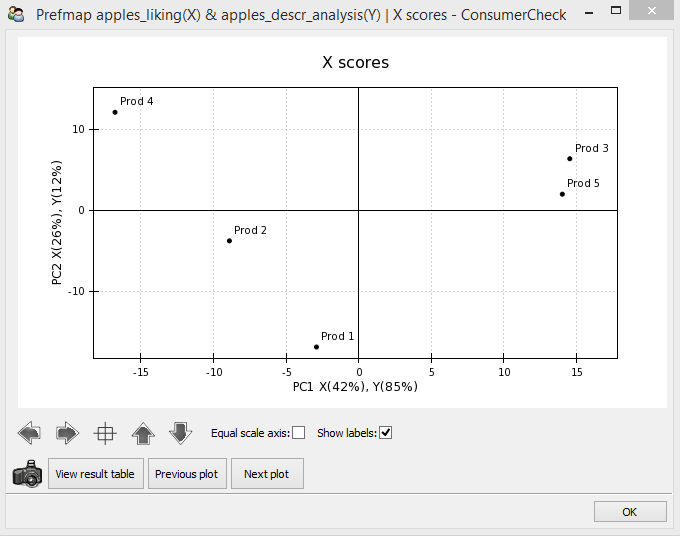}\\
  \caption{A screenshot showing the $X$ scores plot in preference mapping for the \textit{apple consumer liking} and \textit{descriptive analysis / sensory profiling data}.}\label{fig:GUI_prefmap_Xscores}
\end{figure}

The results presented in this section were computed with the following settings: internal preference mapping (i.e. the \textit{consumer liking} data are set to be $X$ in the model and \textit{descriptive analysis / sensory profiling} data are set to be Y); PLSR; $X$ and $Y$ are not standardised since all variables in the respective matrices are based on the same scale. Figure~\ref{fig:GUI_prefmap_Xscores} shows the $X$ scores of the preference mapping model visualising how the products relate to each other in the space spanned by the first two components. As with the PCA scores plot in Figure~\ref{fig:GUI_PCA_scores} similar products are located close to each other and dissimilar products have a larger distance between them. This time, however, the distribution of the products is influenced by the common variance in both $X$ and $Y$ matrix, since in this case PLSR was chosen to compute the preference model. Remember that if PCR is chosen for model computation instead of PLSR, the $X$ scores are computed from matrix $X$ only, while matrix $Y$ has no influence. For details on differences between PLSR and PCR the reader is suggested to consult Section~\ref{sec:statMethod_PLSR} and ~\ref{sec:statMethod_PCR}. \\
As can be seen from Figure~\ref{fig:GUI_prefmap_Xscores} product 3 and 5 are again located close to one another. Products 1, 2 and 4 are again on the opposite side of product 3 and 5 with regard to PC1, however, they are more scattered than they were as with PCA where analysis was based on \textit{descriptive analysis / sensory profiling} data only (Section~\ref{sec:GUI_PCA}). It is important to note how much of the variance in $X$ and $Y$ the first two principal components explain. PC1 and PC2 explain 42\% and 26\% (first number in parenthesis) of the variance in the $X$ matrix. This totals to 68\% for $X$ which is considerable taking into account how noisy \textit{consumer liking} data often can be. PC1 and PC2 explain 85\% and 12\% (second number in parenthesis) of the data in $Y$ (the \textit{descriptive analysis / sensory profiling} data in our case) totalling 97\%. The high levels of explained variance for $X$ and $Y$ indicate that there is a lot of common systematic variation in the data.

\subsubsection[Prefmap - XY correlation loadings]{Preference mapping - $X$\&$Y$ correlation loadings}\label{sec:GUI_PrefMap_XYcorrLoadings}
\begin{figure}
  \centering
  \includegraphics[scale=0.6]{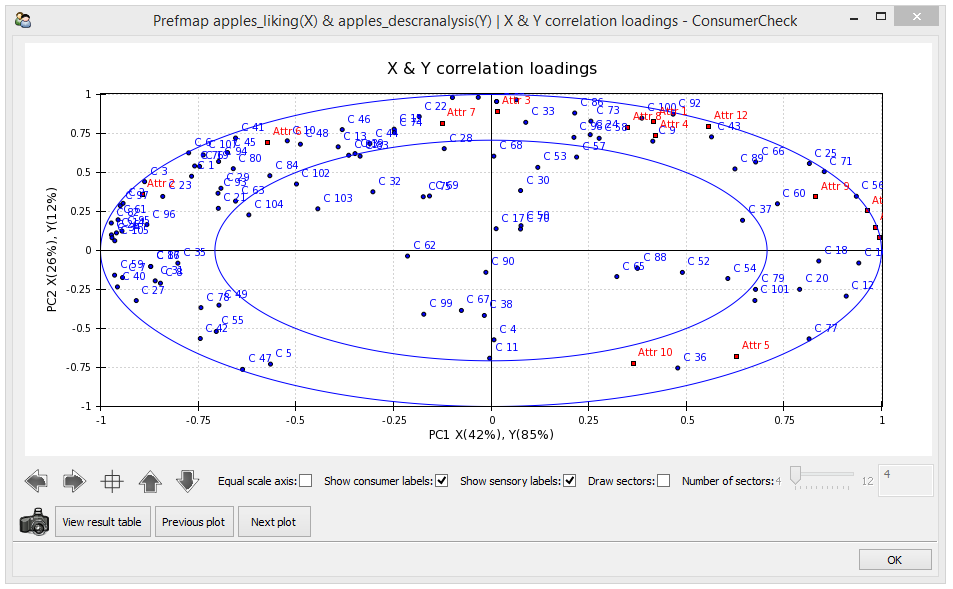}\\
  \caption{A screenshot of the $X$ \& $Y$ correlation loadings plot, the actual preference map for the \textit{apple consumer liking} and \textit{descriptive analysis / sensory profiling data}.}\label{fig:GUI_prefmap_XYcorrLoadings}
\end{figure}

Figure~\ref{fig:GUI_prefmap_XYcorrLoadings} shows the actual preference map that is used for interpretation and visualisation of consumer preferences and drivers of liking. In this plot both the correlation loadings from $X$ and $Y$ are displayed in the same plot.

\begin{figure}
  \centering
  \includegraphics[scale=0.6]{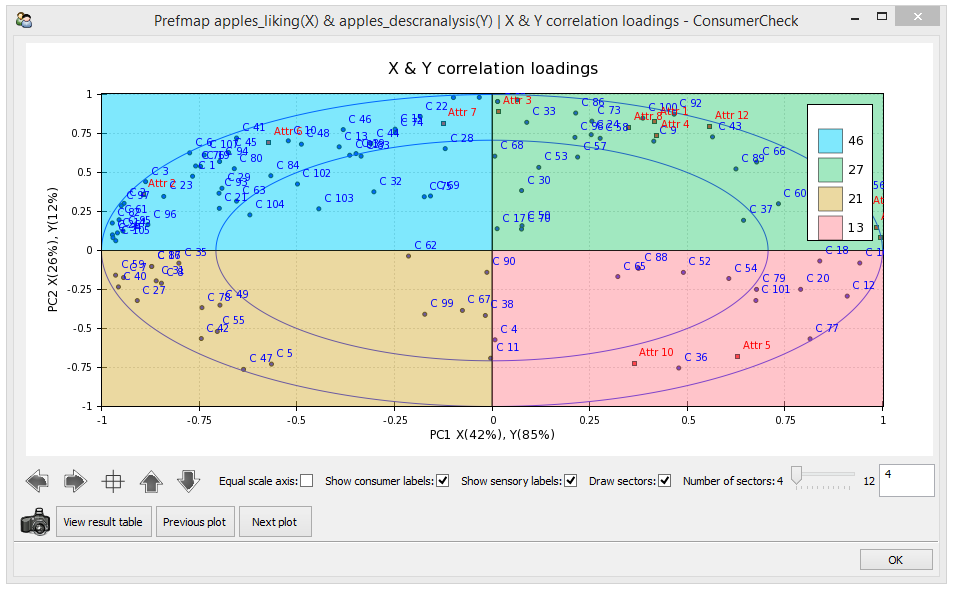}\\
  \caption{A screenshot of the $X$ \& $Y$ correlation loadings plot, the actual preference map for the \textit{apple consumer liking} and \textit{descriptive analysis / sensory profiling data}. This is the same plot as shown in Figure~\ref{fig:GUI_prefmap_XYcorrLoadings}, however here the $X$ \& $Y$ correlation loadings are devided into 4 segments.}\label{fig:GUI_prefmap_XYcorrLoadings_segmented}
\end{figure}

Correlation loadings belonging to matrix $X$ are always coloured in blue. In this example they start with the letter 'C' followed by a number that identifies the consumers that participated in the trial. The correlation loadings from matrix $Y$ are always coloured in red. In this case they are the sensory attributes that describe the product. What we can conclude from Figure~\ref{fig:GUI_prefmap_XYcorrLoadings} is that many consumers prefer products with high intensities of attribute 2 and 6 (upper left part of the plot) since a large part of the consumers are in proximity of those attributes. Attributes 11, 13 and 14, which are all highly correlated, are less preferred although there are a few consumers that prefer high intensities of these sensory attributes. All of them have high explained variances, since they are located very close to the outer ring that indicates 100\% explained variance. Attributes 5 and 10 are also correlated, however to a lesser degree. The explained variances for those two attributes are somewhat lower. Remember that the inner ring indicates 50\% explained variance. Consumers in the inner circle closer to the origo don't discriminate between the products with regard to the variation described by PC1 and PC2. Since the $X$ \& $Y$ correlation loadings plot often is crowded it may be helpful to remove the consumer or sensory attribute labels in order to get a less distorted picture of where consumers and products are located in the plot. This can be done by checking/unchecking the respective boxes ("Show consumer labels", "Show sensory attribute labels") at the bottom of the plot. Furthermore, the $X$ \& $Y$ correlation loadings plot can be divided into segments when checking the "Draw sectors" checkbox. This may be a handy tool to identify quickly which products and attributes are most preferred, that is which products and attributes have most consumers in their proximity in the plot (when correlation loadings plot and $X$ scores are superimposed). By default four segments are drawn as shown in Figure~\ref{fig:GUI_prefmap_XYcorrLoadings_segmented}. The number of segments may be changed by either moving the slider located to the right of the checkbox or by entering the number of segments in the text box to the far right.

As can be seen most consumers are found in the upper left segment which is coloured in blue. The legend indicates that the number of consumers in this segment is 46. The segment with the fewest consumers is found in the lower right corner coloured in pink containing only 13 consumers.

\subsubsection[Prefmap X loadings]{Preference mapping - $X$ loadings}\label{sec:GUI_PrefMap_Xloadings}
\begin{figure}
  \centering
  \includegraphics[scale=0.6]{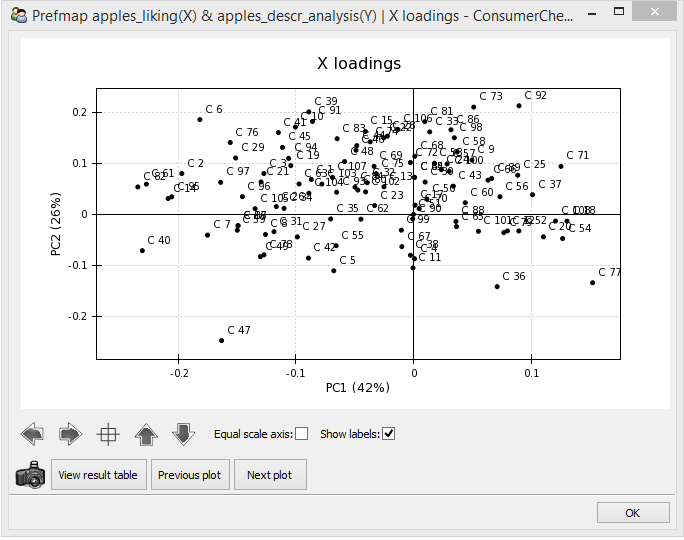}\\
  \caption{A screenshot of the $X$ loadings in preference mapping based on the \textit{apple consumer liking} data.}\label{fig:GUI_prefmap_Xloadings}
\end{figure}

The $X$ loadings in preference mapping show how the variables of the $X$ matrix contribute to the common variation between $X$ and $Y$ for each principal component. Figure~\ref{fig:GUI_prefmap_Xloadings} shows an example of $X$ loadings for PC1 and PC2. As mentioned previously, \textit{consumer liking} data were chosen to be the $X$ matrix in the model, hence the variables of the $X$ matrix are consumers that have tested the products. In Figure~\ref{fig:GUI_prefmap_Xloadings} we can see how consumers spread out across the plane spanned by PC1 and PC2, providing information on how much variance every consumer contributed to the variance explained by PC1 and PC2.

\subsubsection[Prefmap Y loadings]{Preference mapping - $Y$ loadings}\label{sec:GUI_PrefMap_Yloadings}
\begin{figure}
  \centering
  \includegraphics[scale=0.6]{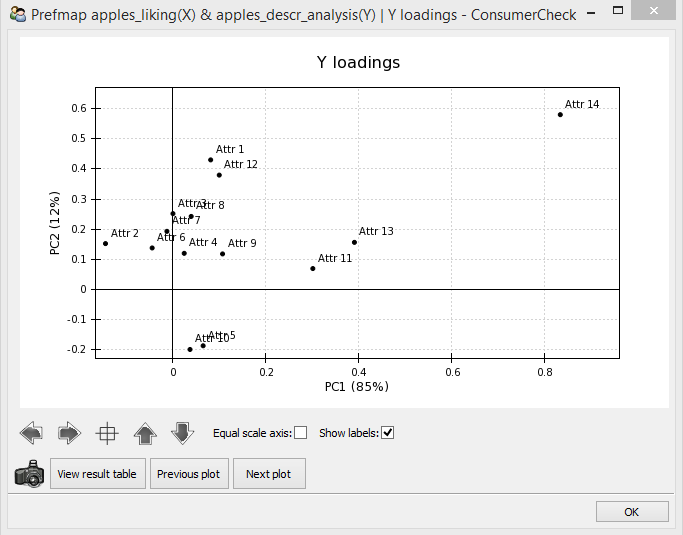}\\
  \caption{A screenshot of the $Y$ loadings in preference mapping based on the \textit{apple descriptive analysis / sensory profiling} data.}\label{fig:GUI_prefmap_Yloadings}
\end{figure}

The $Y$ loadings in preference mapping show how the variables of the $Y$ matrix contribute to the common variation between $X$ and $Y$ for each principal component. Note that PLSR was used for the computation of this preference model and that the statement above is true only for PLSR, not PCR. This is because with PCR the components are determined by the $X$ matrix only and the variables in $Y$ are projected subsequently onto the model (see sections \ref{sec:statMethod_PLSR} and \ref{sec:statMethod_PCR}). Figure~\ref{fig:GUI_prefmap_Yloadings} shows an example of $Y$ loadings for PC1 and PC2 visualising how much variation each variable contributes to the explained variance described by PC1 and PC2.

\subsubsection[Prefmap explained variances X]{Preference mapping - explained variances in X}\label{sec:GUI_PrefMap_XexplVar}
\begin{figure}
  \centering
  \includegraphics[scale=0.6]{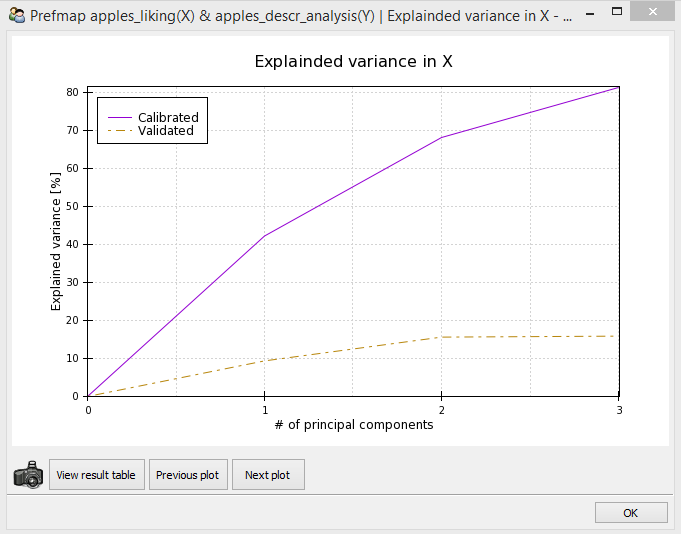}\\
  \caption{A screenshot of the cumulative explained variances in $X$ for the \textit{apple consumer liking} data in preference mapping model.}\label{fig:GUI_prefmap_XexplVar}
\end{figure}

Figure~\ref{fig:GUI_prefmap_XexplVar} shows the cumulative calibrated and validated explained variances for the $X$ matrix. One can see that the calibrated explained variance increases to about 80\% with the first three PC's. The validated explained variance, however, reaches only a level of about 17\%. Low validated explained variances are quite common for \textit{consumer liking} data since these data often are relatively noisy due to the individual differences between consumers and also because of the low number of objects or products in the data. When validating the model with cross validation, the model may change relatively much with each validation step, which then leads to poor predictions of the products left out in the cross validation process. The numerical results can be viewed by clicking on the \textit{View result table} button which opens a new data window displaying the numbers. If needed, the user may select all or parts of the data and use the \textit{copy to clipboard} button at the bottom of the data window to copy and paste the data to another software.

\subsubsection[Prefmap explained variances Y]{Preference mapping - explained variances in Y}\label{sec:GUI_PrefMap_YexplVar}
\begin{figure}
  \centering
  \includegraphics[scale=0.6]{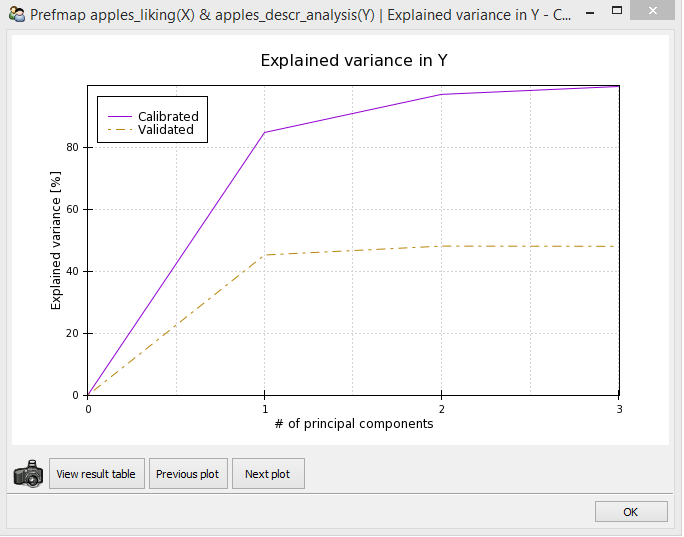}\\
  \caption{A screenshot of the cumulative explained variances in $Y$ \textit{apple descriptive analysis / sensory profiling} data in the preference mapping model.}\label{fig:GUI_prefmap_YexplVar}
\end{figure}

Figure~\ref{fig:GUI_prefmap_YexplVar} shows the cumulative calibrated and validated explained variances for the $Y$ matrix. Here, the calibrated explained variance jumps up to about 85\% and approaches 100\% with the first two PC's. This is quite common for \textit{descriptive analysis / sensory profiling} data, since trained panels typically produce more systematic data than untrained consumers. Numerical results can be accessed by clicking on the \textit{View result table} button. A new data window appears then where numbers are viewed and may be copied to other softwares.

\subsection[PLSR/PCR tab]{Partial least squares regression and principal component regression}\label{sec:GUI_PLSR_PCR}
\begin{figure}
  \centering
  \includegraphics[scale=0.6]{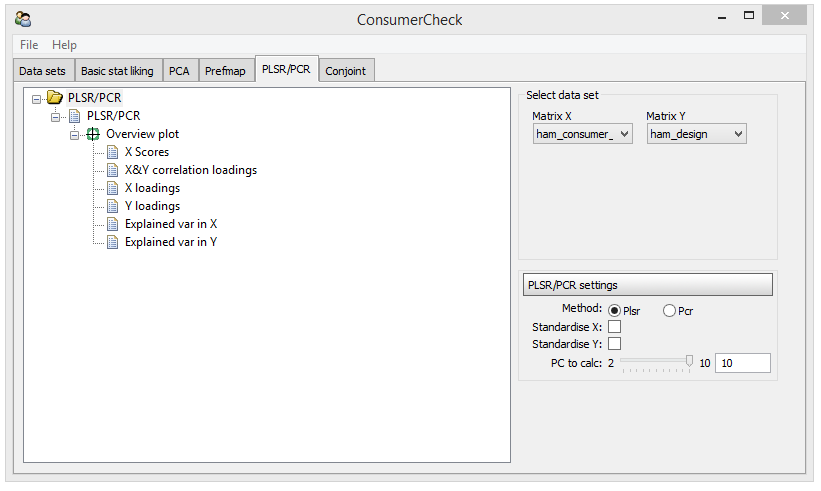}\\
  \caption{A screenshot of the \textit{PLSR/PCR} tab.}\label{fig:GUI_PLSR-PCR}
\end{figure}

The implementation of the \textit{PLSR/PCR} tab is actually almost identical to the \textit{preference mapping} tab that was described above (see Section~\ref{sec:GUI_PrefMap}). The most important difference is that the \textit{PLSR/PCR} tab does not restrict its use to only \textit{consumer liking} and \textit{descriptive analysis / sensory profiling} data, but also allows for analysis of data tagged as \textit{consumer characteristics}, \textit{design matrix} and \textit{other}. Another difference is that the \textit{PLSR/PCR} tab doesn't provide the functionality for segmentation in the $X$ \& $Y$ correlation loadings plot. Furthermore, there are some minor differences regarding the structure of the GUI when compared to the GUI of the \textit{Prefmap} tab (compare Figure~\ref{fig:GUI_prefmap} and \ref{fig:GUI_PLSR-PCR}). In the upper right corner of the \textit{PLSR/PCR} tab the user can choose with the two drop down menus which data is set to be $X$ and $Y$ in the statistical model. Recall that in the \textit{Prefmap} tab at the same place the user is supposed to set the \textit{consumer liking} data and \textit{descriptive analysis / sensory profiling} data for which the \textit{preference mapping} model will be computed for. The other difference between the respective GUI's is that there are no so-called radio-buttons in the \textit{PLSR/PCR} tab for the selection of \textit{internal} or \textit{external preference mapping}, since here the multivariate regression model now is of general character, not focused on only \textit{consumer liking} data and \textit{descriptive analysis / sensory profiling} data.

Other than that the concept of computation and presentation of $X$ scores, $X$ \& $Y$ correlation loadings, $X$ loadings, $Y$ loadings, explained variance in $X$ and $Y$ is identical to those in the \textit{Prefmap} tab and will not be repeated in this section.

\subsection[Conjoint analysis tab]{Conjoint analysis}\label{sec:GUI_Conjoint}
\begin{figure}
  \centering
  \includegraphics[width = \textwidth]{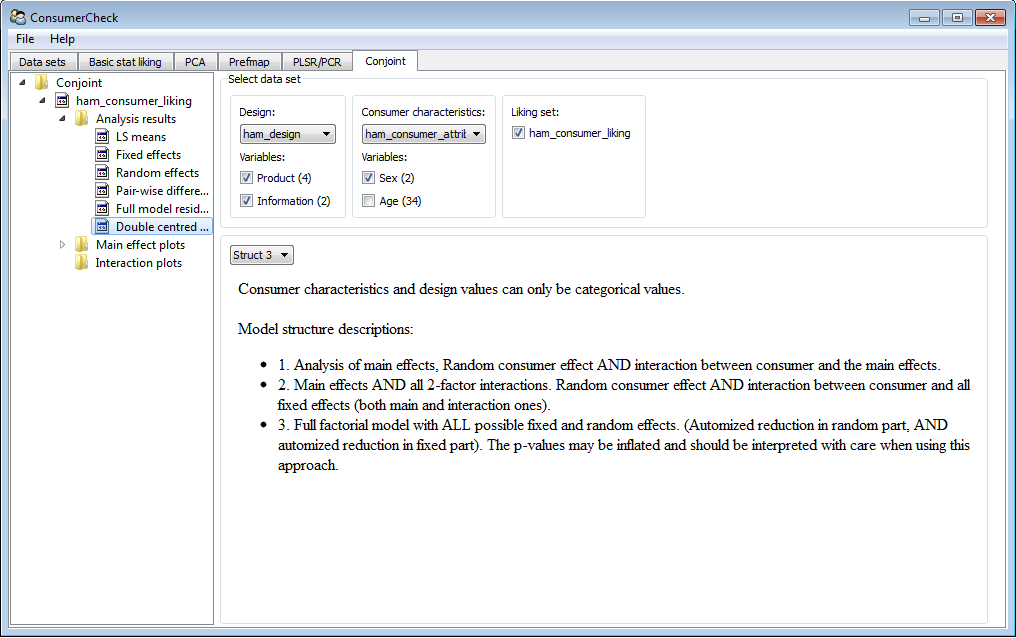}\\
  \caption{A screenshot of the GUI for conjoint analysis where the \textit{ham product design} data, \textit{ham consumer liking} data and \textit{ham consumer characteristics} data are selected for analysis.}\label{fig:GUI_conjoint}
\end{figure}

In order to run a conjoint analysis \textit{product design} data, \textit{consumer liking} data and \textit{consumer characteristics} data are required. Figure~\ref{fig:GUI_conjoint} shows the screenshot of the conjoint GUI. As usual, model settings may be set with the controls on the right side of the GUI. The \textit{product design} data and \textit{consumer characteristics} data that are used for computation of the conjoint model are selected from their respective drop-down menus. As soon as the data is selected in the drop-down menu its variables are displayed as checkboxes right below. Data of type \textit{consumer liking} appear to the right of the two drop-down menus and may be selected with checkboxes. The reason why a drop-down menu wasn't used for these data is that in some consumer trials one may have asked the consumers to rate their liking on the same set of products for multiple modalities, such as odour, flavour, texture, etc. Having checkboxes rather than one drop-down menu ConsumerCheck allows the user to select multiple \textit{consumer liking} data and to compute multiple conjoint models, i.e. one for each modality. By doing so, this generates multiple tree controls, one for each conjoint model, on the left side of the GUI providing access to multiple conjoint models at once. This may be convenient if the user wants to jump quickly between the models and compare effects of design variables and consumer characteristics for the different liking data. Below the tools for setting the data for conjoint analysis there is another drop-down menu where the user can select the complexity of the conjoint model. A short description of complexity for each conjoint model structure is given below the drop-down menu. In this paper a more detailed description of each conjoint model structure is provided:

\begin{description}
\item[Struct 1] 	The mixed effects model includes fixed main effects. Random effects consist of random consumer effect and interaction between consumer and the main effects.
\item[Struct 2] 	The mixed effects model includes main effects and all 2-factor interactions. Random effects consist of consumer effect and interaction between consumer and all fixed effects (both main and interaction ones).
\item[Struct 3] 	This is a full factorial model with all possible fixed and random effects (i.e. including all main effects and all higher-way interactions). The automated reduction in random part is followed by an automated reduction in fixed part. The tests for the random effects use likelihood ratio tests while the tests for the fixed effects use the F-test with Satterthwaite's approximation to degrees of freedom. The automated reduction in the fixed part uses the principle of marginality, i.e. the highest order interactions are tested first: if they are significant, the lower order effects are not eliminated even if being non-significant. This type of structure uses the methodology from \citet{kuznetsova2015a}.
\end{description}

There are multiple items at the tree-control from which computational result may be accessed, either in numeric format in tables or as plots. Each of the results provided are described in more detail below. For an illustration the ham data (Figure~\ref{fig:GUI_conjoint}) were analysed using the conjoint method. The following factors were selected for the model, i.e. they were checked below the drop-down menus: 'product' and 'information' from the \textit{design matrix} as well as 'sex' from the \textit{consumer characteristics}. Age was not included in model since it has too many levels (34), which could cause long computation times or freezing of the software. To avoid this problem, one create a new variable where the age of the consumers is divided into several bins or categories, as for example: 1 when age is between 15 and 29; 2 when age is between 30 and 49; etc and include this in the conjoint model. Finally, \textit{Struct 2} was selected, which means that all selected factors and all their 2-factor interactions were included in the model.

\subsubsection[LS means]{Conjoint analysis - LS means}\label{sec:GUI_Conjoint_lsmeans}
The LS means Table~\ref{tab:GUI_Conjoint_lsmeans} shows population means. In case of balanced data they are exactly the corresponding means. From the column named "Estimate" one may see that the most liked products are Product 3 and 4. Standard errors for the population means, lower and upper 95 percents confidence intervals are also provided.

\begin{table}[h]
\caption{\textit{LS means} result table.} \label{tab:GUI_Conjoint_lsmeans}
\resizebox{\textwidth}{!}{%
\begin{tabular}{llllllllll}
\hline
Model parameter          & Information & Product & Sex & Estimate & Standard Error & DF    & t-value & Lower CI & Upper CI \\ \hline
Information  1           & 1           & NA      & NA  & 5.6318   & 0.1414         & 106.2 & 39.83   & 5.3514   & 5.9121   \\
Information  2           & 2           & NA      & NA  & 5.8312   & 0.1414         & 106.2 & 41.24   & 5.5509   & 6.1116   \\
Product 1                & NA          & 1       & NA  & 5.8084   & 0.233          & 309.5 & 24.93   & 5.3499   & 6.2668   \\
Product 2                & NA          & 2       & NA  & 5.1012   & 0.233          & 309.5 & 21.89   & 4.6428   & 5.5597   \\
Product 3                & NA          & 3       & NA  & 6.0909   & 0.233          & 309.5 & 26.14   & 5.6324   & 6.5493   \\
Product 4                & NA          & 4       & NA  & 5.9256   & 0.233          & 309.5 & 25.43   & 5.4672   & 6.384    \\
Sex 1                    & NA          & NA      & 1   & 5.8537   & 0.1831         & 79    & 31.97   & 5.4892   & 6.2181   \\
Sex 2                    & NA          & NA      & 2   & 5.6094   & 0.1854         & 79    & 30.26   & 5.2404   & 5.9784   \\
Information:Product  1 1 & 1           & 1       & NA  & 5.7287   & 0.2541         & 423.2 & 22.54   & 5.2292   & 6.2282   \\
Information:Product  2 1 & 2           & 1       & NA  & 5.8881   & 0.2541         & 423.2 & 23.17   & 5.3885   & 6.3876   \\
Information:Product  1 2 & 1           & 2       & NA  & 4.8981   & 0.2541         & 423.2 & 19.27   & 4.3986   & 5.3976   \\
Information:Product  2 2 & 2           & 2       & NA  & 5.3043   & 0.2541         & 423.2 & 20.87   & 4.8048   & 5.8039   \\
Information:Product  1 3 & 1           & 3       & NA  & 5.8754   & 0.2541         & 423.2 & 23.12   & 5.3758   & 6.3749   \\
Information:Product  2 3 & 2           & 3       & NA  & 6.3063   & 0.2541         & 423.2 & 24.81   & 5.8068   & 6.8059   \\
Information:Product  1 4 & 1           & 4       & NA  & 6.025    & 0.2541         & 423.2 & 23.71   & 5.5254   & 6.5245   \\
Information:Product  2 4 & 2           & 4       & NA  & 5.8263   & 0.2541         & 423.2 & 22.93   & 5.3267   & 6.3258   \\
Information:Sex  1 1     & 1           & NA      & 1   & 5.7073   & 0.1987         & 106.2 & 28.72   & 5.3133   & 6.1013   \\
Information:Sex  2 1     & 2           & NA      & 1   & 6        & 0.1987         & 106.2 & 30.19   & 5.606    & 6.394    \\
Information:Sex  1 2     & 1           & NA      & 2   & 5.5563   & 0.2012         & 106.2 & 27.61   & 5.1573   & 5.9552   \\
Information:Sex  2 2     & 2           & NA      & 2   & 5.6625   & 0.2012         & 106.2 & 28.14   & 5.2636   & 6.0614   \\
Product:Sex  1 1         & NA          & 1       & 1   & 5.8293   & 0.3275         & 309.5 & 17.8    & 5.185    & 6.4736   \\
Product:Sex  2 1         & NA          & 2       & 1   & 5.4024   & 0.3275         & 309.5 & 16.5    & 4.7581   & 6.0468   \\
Product:Sex  3 1         & NA          & 3       & 1   & 6.2317   & 0.3275         & 309.5 & 19.03   & 5.5874   & 6.876    \\
Product:Sex  4 1         & NA          & 4       & 1   & 5.9512   & 0.3275         & 309.5 & 18.17   & 5.3069   & 6.5955   \\
Product:Sex  1 2         & NA          & 1       & 2   & 5.7875   & 0.3315         & 309.5 & 17.46   & 5.1352   & 6.4398   \\
Product:Sex  2 2         & NA          & 2       & 2   & 4.8      & 0.3315         & 309.5 & 14.48   & 4.1477   & 5.4523   \\
Product:Sex  3 2         & NA          & 3       & 2   & 5.95     & 0.3315         & 309.5 & 17.95   & 5.2977   & 6.6023   \\
Product:Sex  4 2         & NA          & 4       & 2   & 5.9      & 0.3315         & 309.5 & 17.8    & 5.2477   & 6.5523   \\ \hline
\end{tabular}}
\end{table}

\subsubsection[Fixed effects]{Conjoint analysis - Fixed effects}\label{sec:GUI_Conjoint_fixed}
Table~\ref{tab:GUI_Conjoint_fixed} shows the marginal ANOVA table for fixed effects. Since in this example \textit{Struct 2} was selected, no reduction of the fixed effects was performed. For \textit{Struct 3} the elimination of non-significant effects is performed and an additional column named "elim.num" is provided that shows the order of elimination of effects. From the table it is seen that only main effects for 'Product' and 'Information' seem to be significant: the $p$~value for the 'Product' effect is around $0.01$, the $p$~value for the 'Information' effect is slightly higher than $0.05$.


\begin{table}[]
\centering
\caption{\textit{Fixed effects} result table.}
\label{tab:GUI_Conjoint_fixed}
\begin{tabular}{lllllll}
\hline
Model parameters    & Sum Sq   & Mean Sq  & NumDF & DenDF    & F.value  & Pr(>F) \\ \hline
Information         & 5.237084 & 5.237084 & 1     & 78.9678  & 3.29105  & 0.073             \\
Product             & 17.92103 & 5.973675 & 3     & 236.9828 & 3.819272 & 0.011             \\
Sex                 & 1.382455 & 1.382455 & 1     & 78.98244 & 0.87894  & 0.351             \\
Information:Product & 10.38735 & 3.462449 & 3     & 239.98   & 2.201362 & 0.089             \\
Information:Sex     & 1.130503 & 1.130503 & 1     & 78.9678  & 0.718754 & 0.399             \\
Product:Sex         & 1.64384  & 0.547947 & 3     & 236.9828 & 0.348374 & 0.79              \\ \hline
\end{tabular}
\end{table}

\subsubsection[Random effects]{Conjoint analysis - Random effects}\label{sec:GUI_Conjoint_random}
Table~\ref{tab:GUI_Conjoint_random} shows an ANOVA-like table for the random effects. Here each random effect was tested with likelihood ratio test. Non-significant random effects were sequentially eliminated if being non-significant according to the default Type 1 error rate 0.1. From the table it is seen that  the effect corresponding to interaction between Product and Consumer is highly significant.


\begin{table}[]
\centering
\caption{\textit{Random effects} result table.} \label{tab:GUI_Conjoint_random}
\begin{tabular}{llll}
\hline
Model paramters      & Chi.sq   & Chi.DF & p.value        \\ \hline
Information:Consumer & 1.339746 & 1      & 0.247          \\
Product:Consumer     & 167.4856 & 1      & \textless0.001 \\
Consumer             & 2.197325 & 1      & 0.138          \\ \hline
\end{tabular}
\end{table}

\subsubsection[Pairwise differences]{Conjoint analysis - Pairwise differences}\label{sec:GUI_Conjoint_pairwiseDiff}
Table~\ref{tab:GUI_Conjoint_pairwise} shows the first part of pairwise comparisons for the fixed factors from Table~\ref{tab:GUI_Conjoint_fixed}. The last part of the table was omitted because of its length. Column "p-value.adjust" is a $p$-value with Bonferroni multiple testing correction within each effect. From this table it is seen  that e.g. Products 2 and 3 are significantly different from one another.

\begin{table}[h]
\caption{Part of the \textit{Pair-wise differences} result table.} \label{tab:GUI_Conjoint_pairwise}
\resizebox{\textwidth}{!}{%
\begin{tabular}{lllllllll}
\hline
Model parameters               & Estimate & Standard Error & DF    & t-value & Lower CI & Upper CI & p-value & p-value.adjust \\ \hline
Information 1-2                & -0.1995  & 0.11           & 79    & -1.81   & -0.4183  & 0.0194   & 0.0735  & 0.0735         \\
Product 1-2                    & 0.7072   & 0.3154         & 237   & 2.24    & 0.0858   & 1.3286   & 0.0259  & 0.1554         \\
Product 1-3                    & -0.2825  & 0.3154         & 237   & -0.9    & -0.9039  & 0.3389   & 0.3714  & 1              \\
Product 1-4                    & -0.1172  & 0.3154         & 237   & -0.37   & -0.7386  & 0.5042   & 0.7105  & 1              \\
Product 2-3                    & -0.9896  & 0.3154         & 237   & -3.14   & -1.611   & -0.3682  & 0.0019  & 0.0114         \\
Product 2-4                    & -0.8244  & 0.3154         & 237   & -2.61   & -1.4458  & -0.203   & 0.0095  & 0.057          \\
Product 3-4                    & 0.1652   & 0.3154         & 237   & 0.52    & -0.4561  & 0.7866   & 0.6009  & 1              \\
Sex 1-2                        & 0.2443   & 0.2606         & 79    & 0.94    & -0.2744  & 0.7629   & 0.3514  & 0.3514         \\
Information:Product  1 1 - 2 1 & -0.1593  & 0.203          & 315.4 & -0.78   & -0.5588  & 0.2401   & 0.4331  & 1              \\
Information:Product  1 1 - 1 2 & 0.8306   & 0.3448         & 326.2 & 2.41    & 0.1522   & 1.509    & 0.0166  & 0.4648         \\
Information:Product  1 1 - 2 2 & 0.4244   & 0.3483         & 334.3 & 1.22    & -0.2607  & 1.1094   & 0.2239  & 1              \\
Information:Product  1 1 - 1 3 & -0.1467  & 0.3448         & 326.2 & -0.43   & -0.825   & 0.5317   & 0.6709  & 1              \\
Information:Product  1 1 - 2 3 & -0.5776  & 0.3483         & 334.3 & -1.66   & -1.2627  & 0.1075   & 0.0981  & 1              \\
Information:Product  1 1 - 1 4 & -0.2962  & 0.3448         & 326.2 & -0.86   & -0.9746  & 0.3821   & 0.3909  & 1              \\
Information:Product  1 1 - 2 4 & -0.0976  & 0.3483         & 334.3 & -0.28   & -0.7826  & 0.5875   & 0.7796  & 1              \\
Information:Product  2 1 - 1 2 & 0.99     & 0.3483         & 334.3 & 2.84    & 0.3049   & 1.675    & 0.0048  & 0.1344         \\
Information:Product  2 1 - 2 2 & 0.5837   & 0.3448         & 326.2 & 1.69    & -0.0947  & 1.2621   & 0.0915  & 1              \\
Information:Product  2 1 - 1 3 & 0.0127   & 0.3483         & 334.3 & 0.04    & -0.6724  & 0.6977   & 0.971   & 1              \\
Information:Product  2 1 - 2 3 & -0.4183  & 0.3448         & 326.2 & -1.21   & -1.0966  & 0.2601   & 0.226   & 1              \\
Information:Product  2 1 - 1 4 & -0.1369  & 0.3483         & 334.3 & -0.39   & -0.822   & 0.5482   & 0.6945  & 1              \\
Information:Product  2 1 - 2 4 & 0.0618   & 0.3448         & 326.2 & 0.18    & -0.6166  & 0.7402   & 0.8579  & 1              \\
Information:Product  1 2 - 2 2 & -0.4063  & 0.203          & 315.4 & -2      & -0.8057  & -0.0068  & 0.0462  & 1              \\
Information:Product  1 2 - 1 3 & -0.9773  & 0.3448         & 326.2 & -2.83   & -1.6557  & -0.2989  & 0.0049  & 0.1372         \\ \hline
\end{tabular}}
\end{table}

\subsubsection[Main effects plot]{Conjoint analysis - Main effects plot}\label{sec:GUI_Conjoint_mainEffPlot}
\begin{figure}
  \centering
  \includegraphics[scale=0.6]{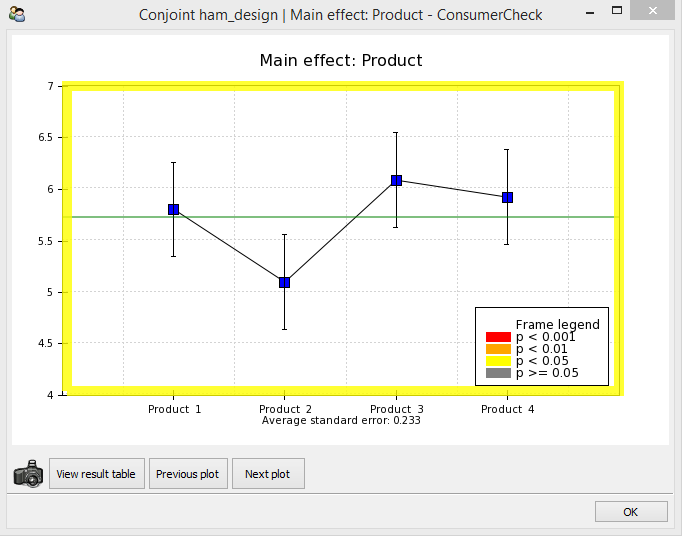}\\
  \caption{Example of a main effect plot for main effect 'Product' from conjoint analysis of the ham data.}\label{fig:GUI_conjoint_mainEffect}
\end{figure}

Figure~\ref{fig:GUI_conjoint_mainEffect} shows population means with their respective 95 percent confidence intervals. From this plot it is seen that Product 3 is the most liked and Product 2 is the least liked.

\subsubsection[Interaction plot]{Conjoint analysis - Interaction plot}\label{sec:GUI_Conjoint_interactionPlot}
\begin{figure}
  \centering
  \includegraphics[scale=0.6]{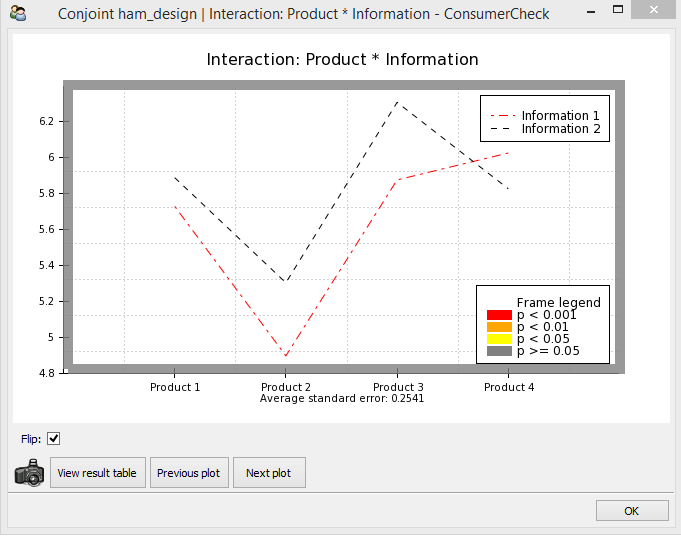}\\
  \caption{Interaction plot for interaction between main effects 'Information' and 'Product' from conjoint analysis of the ham data.}\label{fig:GUI_conjoint_interaction}
\end{figure}

Figure~\ref{fig:GUI_conjoint_interaction} shows a two-way interaction plot (if interaction effects are part of the fixed structure). From this plot users may also observe whether there is an interaction between factors.

\subsection[Individual differences tab]{Individual differences}\label{sec:GUI_individualDifferences}
\begin{figure}
  \centering
  \includegraphics[scale=0.6]{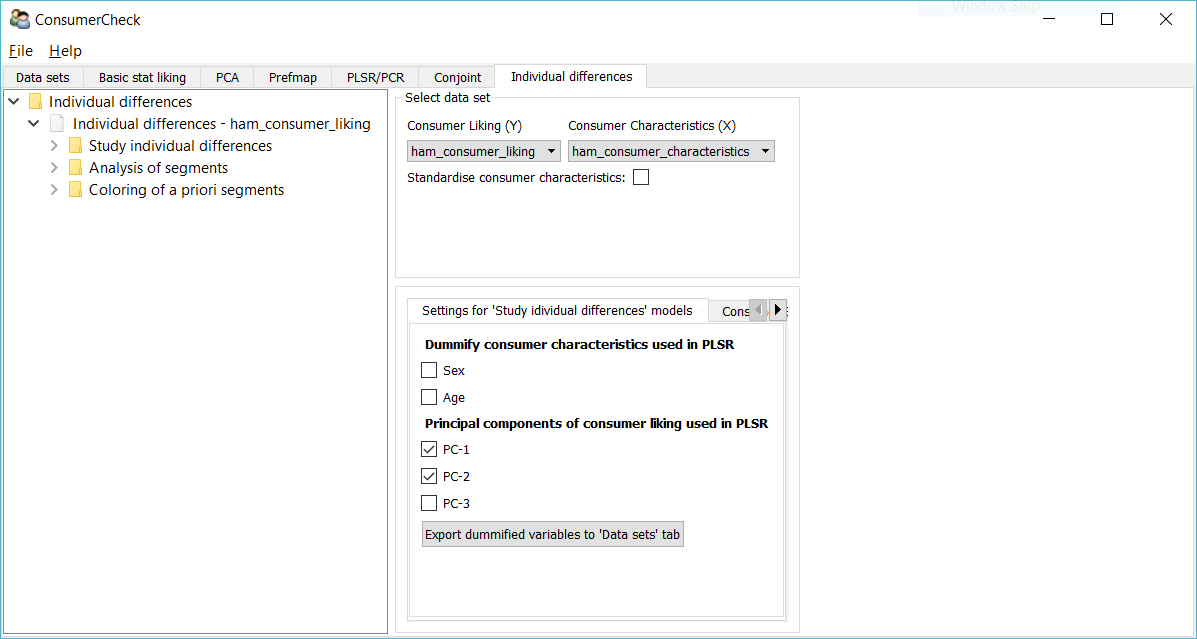}\\
  \caption{A screenshot of the \textit{Individual differences} tab.}\label{fig:GUI_individualDifferences}
\end{figure}

When clicking on the individual differences tab, one first has to define the two data sets to be used from the drop-down menus for \textit{Consumer liking (Y)} data and \textit{Consumer Characteristics (X)} data. Then one can choose between different options as shown in Figure~\ref{fig:GUI_individualDifferences}.

The options available are listed to the left. As can be seen in Figure~\ref{fig:GUI_individualDifferences}, there are three main categories, \textit{Study individual differences}, \textit{Analysis of segments} and \textit{Coloring of a priori segments}.

\begin{figure}
  \centering
  \includegraphics[scale=0.6]{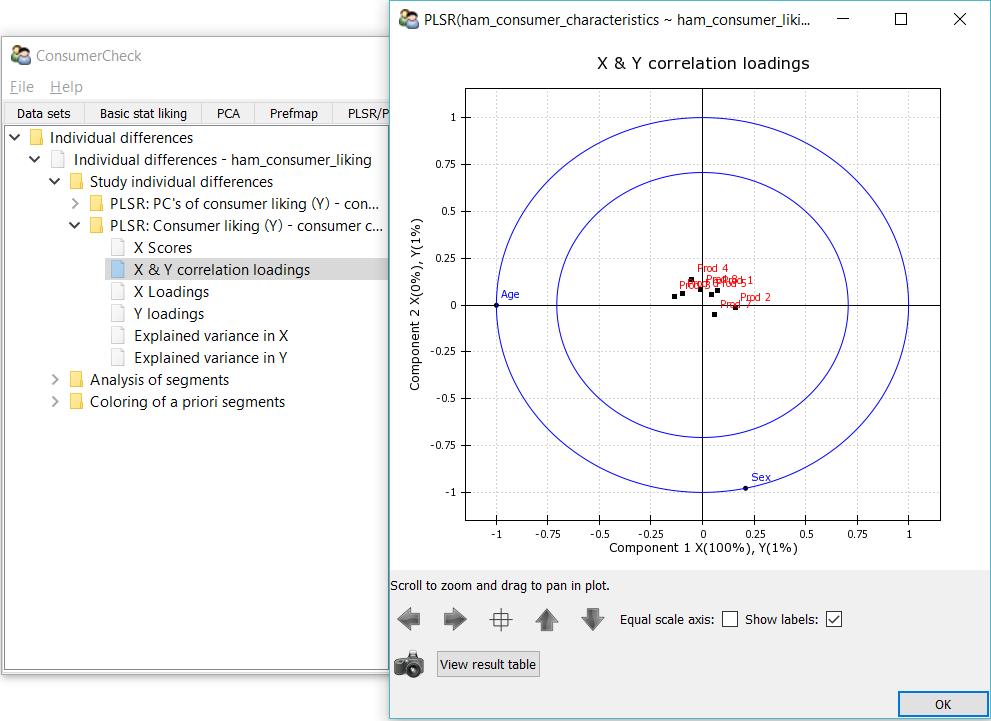}\\
  \caption{Correlation loadings plot based on individual differences in consumer liking and consumer characteristics.}\label{fig:GUI_individualDifferences_X&YcorrLoadings}
\end{figure}

Under the first of these categories, i.e. \textit{Study individual differences}, one can choose between (I) \textit{PLSR: Consumer liking (Y) - consumer characteristics (X)}: a full PLS study of the liking of all samples vs. the individual attributes (Figure~\ref{fig:GUI_individualDifferences_X&YcorrLoadings}); (II) \textit{PLSR: PCs of consumer liking (Y) - consumer characteristics (X)}: the analysis of the loadings of specified principal components vs. the same consumer attributes. In the case shown in Figure~\ref{fig:GUI_individualDifferences}, we can see that the program has detected 2 consumer attributes, Sex and Age, and the first principal component has been selected. The program also allows for categorical variables, but these have to be transformed into dummy variables (0, 1), one for each category using the dummify function. Standardisation of input variables is also possible. Both loadings, correlations loadings and scores are available as shown in Figure~\ref{fig:GUI_individualDifferences_X&YcorrLoadings}. In this case a very low percentage of variation is explained in the liking so little can be said about the relations between liking and the two consumer attributes.

\begin{figure}
  \centering
  \includegraphics[scale=0.6]{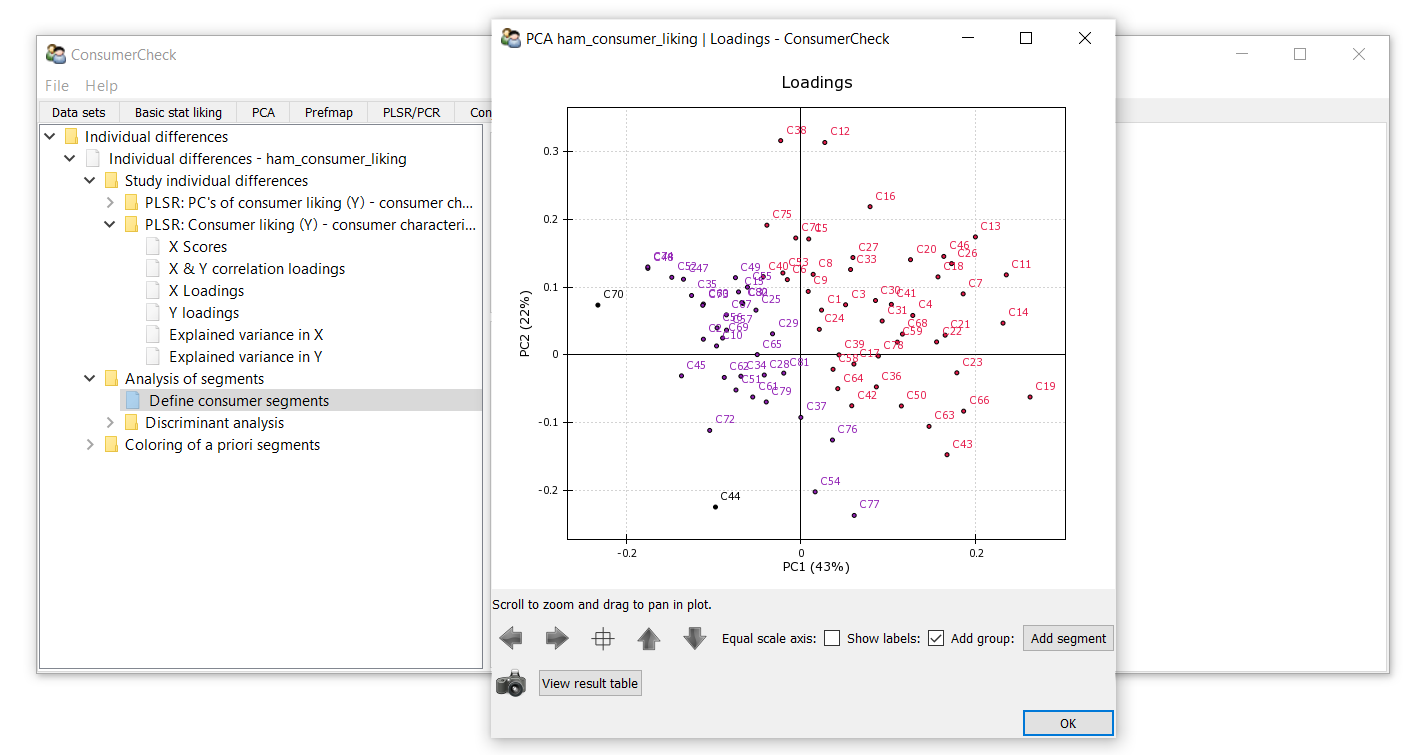}\\
  \caption{In this window the user can define consumer segments for analysis of individual differences.}\label{fig:GUI_individualDifferences_defineConsumerSegments}
\end{figure}

Under the second category, i.e. \textit{Analysis of segments}, one first has to define segments visually by encircling consumers using the cursor as shown in Figure~\ref{fig:GUI_individualDifferences_defineConsumerSegments}. It is important to mention not to move the cursor outside of the frame, which would result in none of the consumers being selected. One needs to click on the button \textit{Add segment} in order to make the marked segment operative. A segment is operative when the labels of the selected consumers change their font colour. One continues until the desired number of segments have been selected. Note that not all samples need to belong to a segment. If two circles overlap, the first selection will be kept. The example in Figure~\ref{fig:GUI_individualDifferences_defineConsumerSegments} shows a situation where some consumers where belong to one out of two segments, as indicated by the two colors purple and red, and some consumers (C44 and C70) are left out of further consideration (black). Here one can see how the different samples relate to the different age groups and genders. In this particular case, the results are not very informative. Invoking the second tab above the specification of dummy variables, one can make the segment data set available and visible under tab \textit{Data sets}.

\begin{figure}
  \centering
  \includegraphics[scale=0.6]{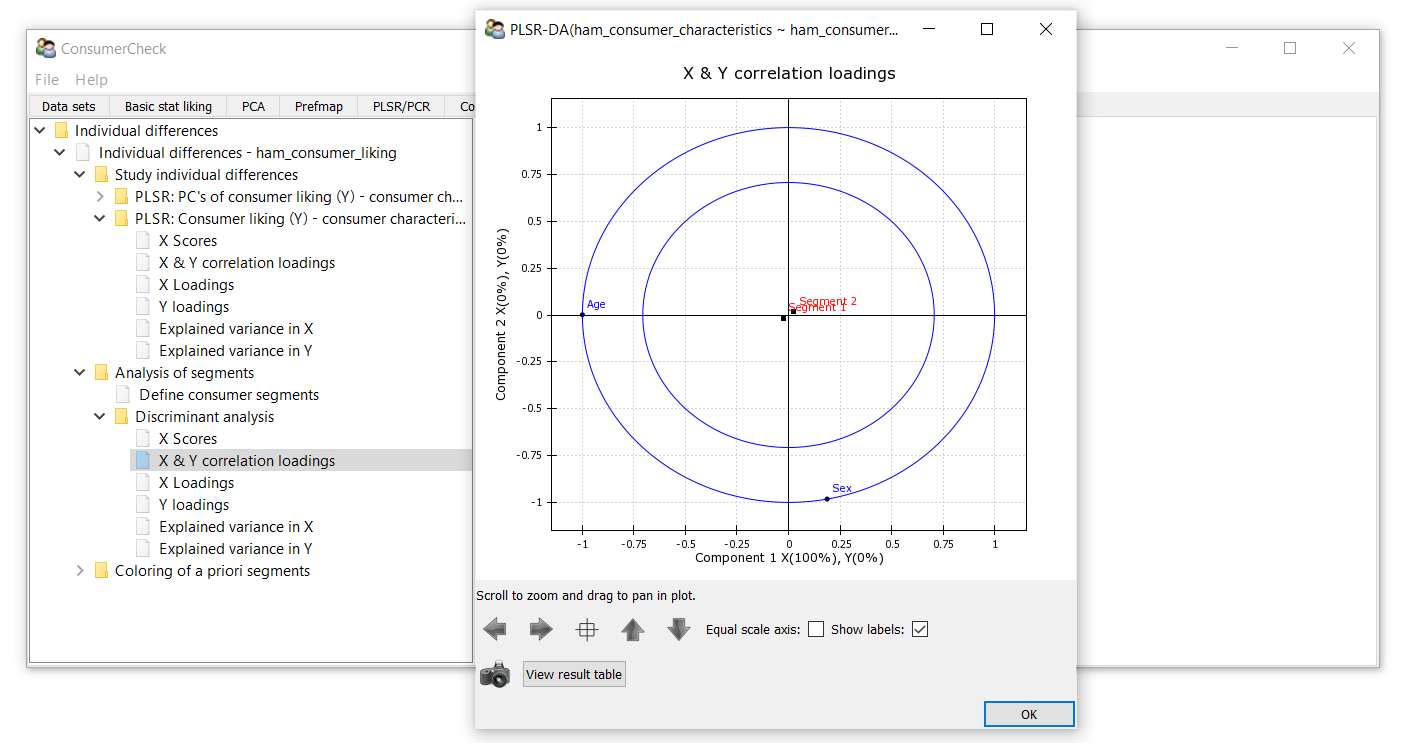}\\
  \caption{Results of PLS-DA based on consumer segments.}\label{fig:GUI_individualDifferences_consumerSegments_PLS-DA}
\end{figure}

As soon as the segments are defined in this way, one can move to the next analysis type \textit{Discriminant analysis}, just below \textit{Define consumer segments}. Here one can generate PLSR plots from PLS-DA based on regressing a matrix of dummy variables representing the segment categories matrix onto the consumer attributes. The results from the segments are presented in Figure ~\ref{fig:GUI_individualDifferences_consumerSegments_PLS-DA}.

As can be noted, the explained variance of Y is zero, implying that the there is no information in the consumer characteristics for prediction of liking.

\section[Conclusion]{Conclusion}\label{sec:conslusion}
ConsumerCheck is an open source data analysis software tailored for analysis of sensory and consumer data. Since some of the implemented methods are generic, such as PCA, PLSR and PCR, other data from other domains may also be analysed with ConsumerCheck. The software comes with a graphical user interface and as such provides non-statisticians and users without programming skills free access to a number of widely used analysis methods within the field of sensory and consumer science. Computational results are presented in plots that are easily generated from the tree-controls within the graphical user interfaces. Since the construction of conjoint analysis models is not always straightforward, ConsumerCheck provides three previously defined model structures of different complexity. ConsumerCheck is an ongoing research project and the objective is to implement further statistical methods over time.

\section[Acknowledgements]{Acknowledgements}\label{sec:acknowledgements}
We would like to thank the Research Council of Norway and Norwegian food industry for funding  the Norwegian part of the ConsumerCheck project. Would also like to thank Direktoratet for F{\o}devareerhverv and the Danish industry for funding the Danish part of the ConsumerCheck project. Help by Kristian Hovde Liland (Norwegian University of Life Sciences) for validation of results computed outside ConsumerCheck, is greatly appreciated.

\bibliographystyle{unsrtnat}


\begin{thebibliography}{30}
  \providecommand{\natexlab}[1]{#1}
  \providecommand{\url}[1]{\texttt{#1}}
  \expandafter\ifx\csname urlstyle\endcsname\relax
    \providecommand{\doi}[1]{doi: #1}\else
    \providecommand{\doi}{doi: \begingroup \urlstyle{rm}\Url}\fi
  
  \bibitem[Lawless and Heymann(2010)]{lawless10}
  Harry~T. Lawless and Hildegarde Heymann.
  \newblock \emph{Sensory Evaluation of Food - Principles and Practices}.
  \newblock Springer, NY, USA, 2nd edition edition, 2010.
  
  \bibitem[N{\ae}s et~al.(2010)N{\ae}s, Brockhoff, and Tomic]{naes10}
  Tormod N{\ae}s, Per~B. Brockhoff, and Oliver Tomic.
  \newblock \emph{Statistics for Sensory and Consumer Science}.
  \newblock Wiley, Chichester, 2010.
  
  \bibitem[Martens and Martens(2001)]{martens01}
  Harald Martens and Magni Martens.
  \newblock \emph{Multivariate analysis of Quality: An Introduction}.
  \newblock Wiley, Chichester, 2001.
  
  \bibitem[Pedregosa et~al.(2011)Pedregosa, Varoquaux, Gramfort, Michel, Thirion,
    Grisel, Blondel, Prettenhofer, Weiss, Dubourg, Vanderplas, Passos,
    Cournapeau, Brucher, Perrot, and Duchesnay]{pedregosa11}
  F.~Pedregosa, G.~Varoquaux, A.~Gramfort, V.~Michel, B.~Thirion, O.~Grisel,
    M.~Blondel, P.~Prettenhofer, R.~Weiss, V.~Dubourg, J.~Vanderplas, A.~Passos,
    D.~Cournapeau, M.~Brucher, M.~Perrot, and E.~Duchesnay.
  \newblock Scikit-learn: Machine learning in {P}ython.
  \newblock \emph{Journal of Machine Learning Research}, 12:\penalty0 2825--2830,
    2011.
  
  \bibitem[Seabold and Perktold(2010)]{seabold2010statsmodels}
  Skipper Seabold and Josef Perktold.
  \newblock statsmodels: Econometric and statistical modeling with python.
  \newblock In \emph{9th Python in Science Conference}, 2010.
  
  \bibitem[Kuznetsova et~al.(2013{\natexlab{a}})Kuznetsova, {Bruun Brockhoff},
    and {Haubo Bojesen Christensen}]{SensMixed}
  Alexandra Kuznetsova, Per {Bruun Brockhoff}, and Rune {Haubo Bojesen
    Christensen}.
  \newblock \textbf{SensMixed}: Mixed effects modelling for sensory and consumer
    data, 2013{\natexlab{a}}.
  \newblock \textbf{R} package version 2.0-5.
  
  \bibitem[Christensen and Brockhoff(2014)]{sensR}
  R.~H.~B. Christensen and P.~B. Brockhoff.
  \newblock \textbf{sensR}---an \textbf{R}-package for sensory discrimination,
    2014.
  \newblock URL \url{http://www.cran.r-project.org/package=sensR/}.
  \newblock \textbf{R} package version 1.4-0.
  
  \bibitem[Husson et~al.(2014)Husson, Le, and Cadoret]{SensoMineR}
  Francois Husson, Sebastien Le, and Marine Cadoret.
  \newblock \textbf{SensoMineR}: Sensory data analysis with \textbf{R}, 2014.
  \newblock URL \url{http://CRAN.R-project.org/package=SensoMineR}.
  \newblock \textbf{R} package version 1.20.
  
  \bibitem[Filzmoser and Varmuza(2015)]{filz15}
  Peter Filzmoser and Kurt Varmuza.
  \newblock \textbf{chemometrics}: Multivariate statistical analysis in
    chemometricschemometrics: Multivariate statistical analysis in chemometrics,
    2015.
  \newblock \textbf{R} package version 1.3.9.
  
  \bibitem[Mevik et~al.(2015)Mevik, Wehrens, and Liland]{mevik15}
  Bj{\o}rn-Helge Mevik, Ron Wehrens, and Kristian~Hovde Liland.
  \newblock \textbf{pls}: Partial least squares and principal component
    regression, 2015.
  \newblock \textbf{R} package version 2.5-0.
  
  \bibitem[Fox and Bouchet-Valat(2015)]{fox15}
  John Fox and Milan Bouchet-Valat.
  \newblock \textbf{Rcmdr}: \textbf{R} commander, 2015.
  \newblock \textbf{R} package version 2.2-3.
  
  \bibitem[Tomic et~al.(2010{\natexlab{a}})Tomic, Risvik, Brockhoff, Nilsen, and
    Naes]{tomic10PCH}
  Oliver Tomic, Henning Risvik, Per~B. Brockhoff, Asgeir Nilsen, and Tormod Naes.
  \newblock Panelcheck, 2010{\natexlab{a}}.
  \newblock URL \url{http://www.panelcheck.com}.
  
  \bibitem[Tomic et~al.(2010{\natexlab{b}})Tomic, Luciano, Nilsen, Hyldig,
    Lorensen, and N{\ae}s]{tomic09}
  Oliver Tomic, Giorgio Luciano, Asgeir Nilsen, Grethe Hyldig, Kirsten Lorensen,
    and Tormod N{\ae}s.
  \newblock Analysing sensory panel performance in a proficiency test using the
    panelcheck software.
  \newblock \emph{European Food Research and Technology}, 230:\penalty0 497--211,
    2010{\natexlab{b}}.
  
  \bibitem[Tomic et~al.(2007)Tomic, Nilsen, Martens, and Naes]{tomic07}
  Oliver Tomic, Asgeir Nilsen, Magni Martens, and Tormod Naes.
  \newblock Visualization of sensory profiling data for performance monitoring.
  \newblock \emph{LWT--Food Science and Technology}, 40:\penalty0 262--269, 2007.
  
  \bibitem[Chavent et~al.(2014)Chavent, Kuentz-Simonet, Labenne, and
    Saracco]{chavent2014}
  Marie Chavent, Vanessa Kuentz-Simonet, Amaury Labenne, and Jerom Saracco.
  \newblock Multivariate analysis of mixed data: The \textbf{PCAmixdata}
    \textbf{R} package.
  \newblock arXiv:1411.4911v3 [stat.CO] 4 Dec 2014, December 2014.
  
  \bibitem[Mardia et~al.(1979)Mardia, Kent, and Bibby]{mardia79}
  K.V. Mardia, J.T. Kent, and J.M. Bibby.
  \newblock \emph{Multivariate Analysis}.
  \newblock London: Academic Press, 1979.
  
  \bibitem[Wold(1982)]{wold82}
  H.~Wold.
  \newblock \emph{Systems under Indirect Observation}, chapter Soft modelling:
    The basics and some extensions.
  \newblock Amsterdam: North Holland, 1982.
  
  \bibitem[Martens and N{\ae}s(1989)]{martens89}
  Harald Martens and Tormod N{\ae}s.
  \newblock \emph{Multivariate Calibration}.
  \newblock John Wiley \& Sons Ltd, Chichester, 1989.
  
  \bibitem[Greenhoff and MacFie(1994)]{green94}
  K.~Greenhoff and H.J.H. MacFie.
  \newblock \emph{Measurements of Food Products}, chapter Preference mapping in
    practice, pages 137--166.
  \newblock Glasgow: Blackie Academic and Professional, 1994.
  
  \bibitem[McEwan(1996)]{mcewan96}
  J.A. McEwan.
  \newblock \emph{Multivariate Analysis of Data in Sensory Science}, volume~16 of
    \emph{Data Handling in Science and Technology}, chapter Preference mapping
    for product optimization, pages 71--102.
  \newblock Amsterdam: Elsevier Science B.V., 1996.
  
  \bibitem[Martens and N{\ae}s(1988)]{martens88}
  Harald Martens and Tormod N{\ae}s.
  \newblock Principal components regression in nir analysis.
  \newblock \emph{Journal of Chemometrics}, 2:\penalty0 155--167, 1988.
  
  \bibitem[Green and Rao(1971)]{green71}
  P.E. Green and V.R. Rao.
  \newblock Conjoint measurement for quantifying judgemental data.
  \newblock \emph{Journal of Marketing Research}, 8:\penalty0 355--363, 1971.
  
  \bibitem[Green and Srinivasan(1978)]{green78}
  P.E. Green and V.~Srinivasan.
  \newblock Conjoint analysis in consumer research: Issues and outlook.
  \newblock \emph{Journal of Consumer Research}, 5:\penalty0 103--123, 1978.
  
  \bibitem[Bates et~al.(2014)Bates, Maechler, Bolker, and Walker]{lme4}
  Douglas Bates, Martin Maechler, Ben Bolker, and Steven Walker.
  \newblock \textbf{lme4}: Linear mixed-effects models using eigen and s4, 2014.
  \newblock URL \url{http://CRAN.R-project.org/package=lme4}.
  \newblock \textbf{R} package version 1.1-7.
  
  \bibitem[Kuznetsova et~al.(2013{\natexlab{b}})Kuznetsova, {Bruun Brockhoff},
    and {Haubo Bojesen Christensen}]{lmerTest}
  Alexandra Kuznetsova, Per {Bruun Brockhoff}, and Rune {Haubo Bojesen
    Christensen}.
  \newblock \textbf{lmerTest}: Tests for random and fixed effects for linear
    mixed effect models (lmer objects of \textbf{lme4} package).,
    2013{\natexlab{b}}.
  \newblock URL \url{http://CRAN.R-project.org/package=lmerTest}.
  \newblock \textbf{R} package version 2.0-12.
  
  \bibitem[N{\ae}s et~al.(2018)N{\ae}s, Varela, and Berget]{naes18}
  Tormod N{\ae}s, Paula Varela, and Ingunn Berget.
  \newblock \emph{Individual differences in sensory and consumer science}.
  \newblock Woodhead Publishing, 2018.
  \newblock ISBN 9780081010006.
  
  \bibitem[Tomic et~al.(2019)Tomic, Graff, Liland, and N{\ae}s]{tomic19}
  Oliver Tomic, Thomas Graff, Kristian~Hovde Liland, and Tormod N{\ae}s.
  \newblock hoggorm: a python library for explorative multivariate statistics.
  \newblock \emph{The Journal of Open Source Software}, 4\penalty0 (39), 2019.
  \newblock \doi{10.21105/joss.00980}.
  \newblock URL \url{http://joss.theoj.org/papers/10.21105/joss.00980}.
  
  \bibitem[Oliphant(2006)]{oliphant06}
  Travis~E Oliphant.
  \newblock \emph{A guide to NumPy}, volume~1.
  \newblock Trelgol Publishing USA, 2006.
  
  \bibitem[Xia et~al.(2010)Xia, McClelland, and Wang]{Xia10}
  Xiao-Qin Xia, Michael McClelland, and Yipeng Wang.
  \newblock Pyper, a python package for using r in python.
  \newblock \emph{Journal of Statistical Software, Code Snippets}, 35\penalty0
    (2):\penalty0 1--8, 7 2010.
  \newblock ISSN 1548-7660.
  \newblock URL \url{http://www.jstatsoft.org/v35/c02}.
  
  \bibitem[Kuznetsova et~al.(2015)Kuznetsova, Christensen, Bavay, and
    Brockhoff]{kuznetsova2015a}
  Alexandra Kuznetsova, Rune~H.B. Christensen, Cecile Bavay, and Per~Bruun
    Brockhoff.
  \newblock Automated mixed anova modeling of sensory and consumer data.
  \newblock \emph{Food Quality and Preference}, 40:\penalty0 31--38, 2015.
  \newblock ISSN 09503293, 18736343.
  \newblock \doi{10.1016/j.foodqual.2014.08.004}.
  
  \end{thebibliography}

\end{document}